\documentclass[aps,prb,twocolumn,showpacs,eqseCNPum,superscriptaddress,floatfix,10pt]{revtex4-1}

\usepackage{amssymb}  
\usepackage{color,soul} 
\usepackage{graphicx}  
\usepackage[tight]{subfigure}  
\usepackage{mathrsfs}
\usepackage[pdftex,pdfusetitle,bookmarks=true,colorlinks=true,citecolor=blue,urlcolor=blue,linkcolor=magenta]{hyperref}
\usepackage{braket}
\usepackage{xfrac}
\usepackage{tikz}
\usepackage{url}
\usepackage{setspace}
\usepackage{enumerate}
\usepackage{bm}
\usepackage{epstopdf}
\usepackage{amsmath}
\usepackage{amsfonts}
\usepackage{mhchem} % for left superscript

\begin{document}

\title{Landau levels in twisted bilayer graphene and semiclassical orbits}

\author{Kasra Hejazi}
\thanks{These two authors contributed equally.}
\affiliation{Department of Physics, University of California Santa Barbara, Santa Barbara, California, 93106, USA}
\author{Chunxiao Liu}
\thanks{These two authors contributed equally.}
\affiliation{Department of Physics, University of California Santa Barbara, Santa Barbara, California, 93106, USA}
\author{Leon Balents}
\affiliation{Kavli Institute for Theoretical Physics, University of California Santa Barbara, CA 93106, USA}

%\date{\today}
\begin{abstract}
Twisted bilayer graphene has been argued theoretically to host exceptionally flat bands when the angle between the two layers falls within a \emph{magic} range near 1.1$^\circ$.  This is now strongly supported by experiment, which furthermore reveals dramatic correlation effects in the magic range due to the relative dominance of interactions when the bandwidth is suppressed.  Experimentally,  quantum oscillations exhibit different Landau level degeneracies when the angles fall in or outside the magic range; these observations can contain crucial information about the low energy physics. In this paper, we report a thorough theoretical study of the Landau level structure of the non-interacting continuum model for twisted bilayer graphene as the magnetic field and the twist angle are tuned.   We first show that a discernible difference exists in the butterfly spectra when twist angle falls in and outside the magic range.   Next, we carry out semiclassical analysis in detail, which quantitatively determines the origin of the low energy Landau levels from the zero field band structure.   We find that the Landau level degeneracy predicted in the above analyses is capable of partially explaining features of the quantum oscillation experiments in a natural way.   Finally, topological aspects, validity, and other subtle points of the model are discussed.
\end{abstract}

\maketitle

\section{Introduction}
In the past year, twisted bilayer graphene (TBG) has attracted immense attention from physicists, following the observation of superconductivity and correlation-induced insulators when the layers are twisted relative to one another close to the ``magic'' angle ($\sim 1.1^\circ$)\cite{cao2018correlated,cao2018unconventional}.  These results have since been confirmed and extended by many independent groups\cite{yankowitz2019tuning,sharpe2019emergent,lu2019superconductors,polshyn2019phonon,cao2019electric,liu2019spin,chen2019signatures}. While there are now many experiments and some results are limited to specific samples at specific angles and densities, so far all indications of correlated behavior have been limited to the density range corresponding to partial fillings of the two low energy bands closest to charge neutrality point (CNP); these two \textit{active} bands are theoretically predicted to show exceptional flatness when the twist angle is tuned to lie within the magic range \cite{Bistritzer2011}.

At the present stage, there is no consensus on the explanation of these effects, despite the many theoretical efforts\cite{xu2018topological,kang2018symmetry,koshino2018maximally,po2018origin,you2018superconductivity,dodaro2018phases,xie2018nature,2018arXiv180704382L,wu2018theory,PhysRevX.8.041041,2019arXiv190108110B}. 
%\blu{[V1: In the theory community, Efforts have been made to approach this problem from the continuum model, tight binding models, first principal calculation, etc.; Electron-electron and/or electron-phonon interactions have been studied; nontrivial topological band structure and topological superconducvitiy have been proposed. ->And add references]} \rd{[V2: In a (large) subset of the theoretic studies, the approach has been to select a basis of non-interacting states corresponding to some number of low energy bands, and then to incorporate electron-electron and/or electron-phonon interactions within it.   Different authors have worked with the Continuum model\cite{koshino2018maximally} (CM) first introduced in Refs.~\onlinecite{Bistritzer2011},\onlinecite{dos2007graphene} and/or tight-binding constructions\cite{kang2018symmetry,xu2018topological} to this end.]} 
A key issue is that the same physics which leads to anomalous narrowing of the low energy bands near the magic angle also makes those bands very sensitive to small details of the model in this regime.  It would be desirable to put direct constraints on the theoretical model from experiment.  

An effective way to obtain an understanding of the low energy degrees of freedom in a two-dimensional system is to probe it with perpendicular magnetic field and study the quantum oscillations; local minima of longitudinal resistivity form straight lines in the plane of carrier density and magnetic field, a fact that can be seen in Landau fan diagrams. Previous experimental results\cite{cao2016superlattice} at the angle $1.8^\circ$ exhibited Landau level (LL) filling factors $\nu = \pm 4, \pm 12, \pm 20, \ldots$ at the CNP (we define $\nu = n \Phi_0/\Phi$, where $n$ is the 2d density measured from charge neutrality, $\Phi_0 =h/e$ is the flux quantum, and $\Phi = B \mathcal{A}$ is the flux per unit cell of the moir\'e pattern, with unit cell area $\mathcal{A}$); note that these numbers double those of monolayer graphene. This can be understood by noting that there are two \textit{renormalized} Dirac points in the moir\'e Brillouin Zone (BZ) where the two active bands touch as predicted by the continuum model\cite{Bistritzer2011,dos2007graphene} (CM) of TBG, and that furthermore one needs to consider $4 = 2\times 2$ copies of the model due to spin and valley degrees of freedom of the electrons. However, study of quantum oscillations in the magic range \cite{yankowitz2019tuning,cao2018unconventional} revealed a different sequence near CNP, namely $\nu = \pm 4, \pm 8, \pm 12, \ldots$. Some authors have argued that this happens due to interactions or some kind of symmetry breaking\cite{po2018origin}.

While certainly the interaction effects are of primary interest, without a firm foundation for the free electron moir\'e physics it is impossible to disambiguate subtle band effects from correlation ones.  Here, inspired by the above observation, we study the effect of perpendicular magnetic field without interactions.  We start with the CM, and incorporate the magnetic field into it; to this end, we use the method introduced in Ref. \onlinecite{bistritzer2011moire} with some modifications. In particular, we have considered the effects of lattice corrugation phenomenologically by differentiating between the tunneling amplitude at AA and AB/BA regions of the moir\'e superlattice \cite{koshino2018maximally}.

We start by first studying angles larger than the magic value, and then restrict our attention to the magic range. We indeed observe that at larger angles, the same sequence mentioned above, $4 \times (\pm1,\pm3,\pm5,\ldots)$, can be seen; however, as the twist angle is reduced into the magic range, we first observe a sequence of $4 \times (\pm 1, \pm 2, \pm 3, \ldots)$, and then a sequence of $4 \times (\pm 1, \pm 4, \ldots)$ upon further decrease of the angle. The former happens close to the twist angle where the $\Gamma$ point (the highest symmetry point) of the BZ becomes gapless in a quadratic band touching and the nonmagnetic active bands become most flat, while the latter happens upon the formation of threefold local minima (maxima) in the upper (lower) active band. Finally we present a semiclassical study at small enough magnetic fields that associates the above results to certain regions in the BZ.

The rest of the paper is organized as follows. In Sec.~\ref{sec:model}, we introduce the magnetic model used here; the details of the model are presented in App.~\ref{app:the_model}. Then in Sec.~\ref{sec:num_sol_semi_classical} the numerical solution of the magnetic model is presented and discussed outside and within the magic range; in particular, relatively small field regime and LL filling factors therein are studied. Furthermore, comparison with results derived from a semiclassical analysis is presented. Finally, in Sec.~\ref{sec:conclusion} these discussions are summarized and also some further results regarding the intermediate and large field regimes, inclusion of particle-hole symmetry breaking terms, etc.~are discussed. Further details of these discussions are presented in the Appendix.

\section{The model }\label{sec:model}

\begin{figure*}[!t]
	\centering
	\subfigure[]{\label{fig:3500}\includegraphics[height=0.485\textwidth]{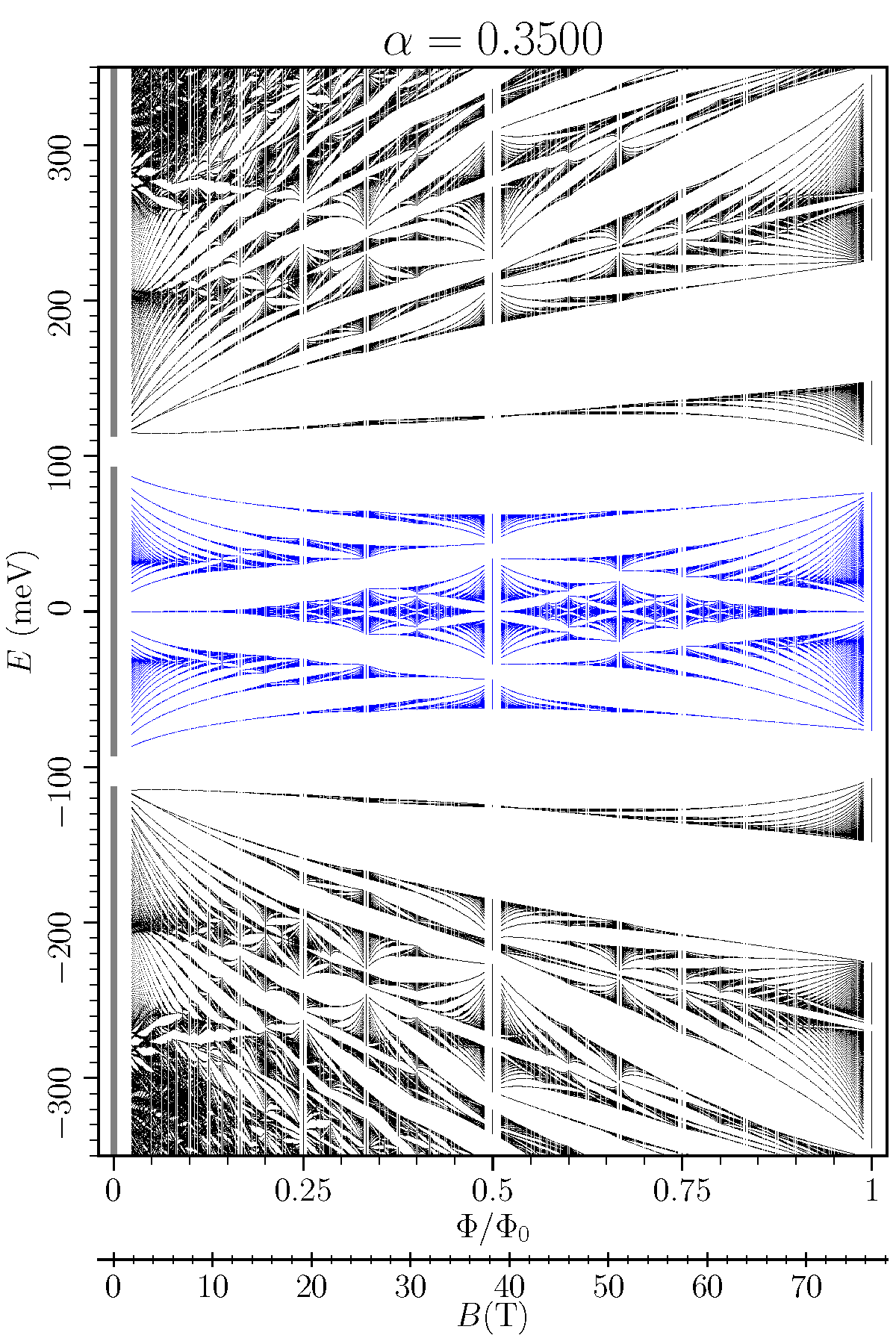}}
	\hspace{-0.1in}
	\subfigure[]{\label{fig:3500}\includegraphics[height=0.485\textwidth]{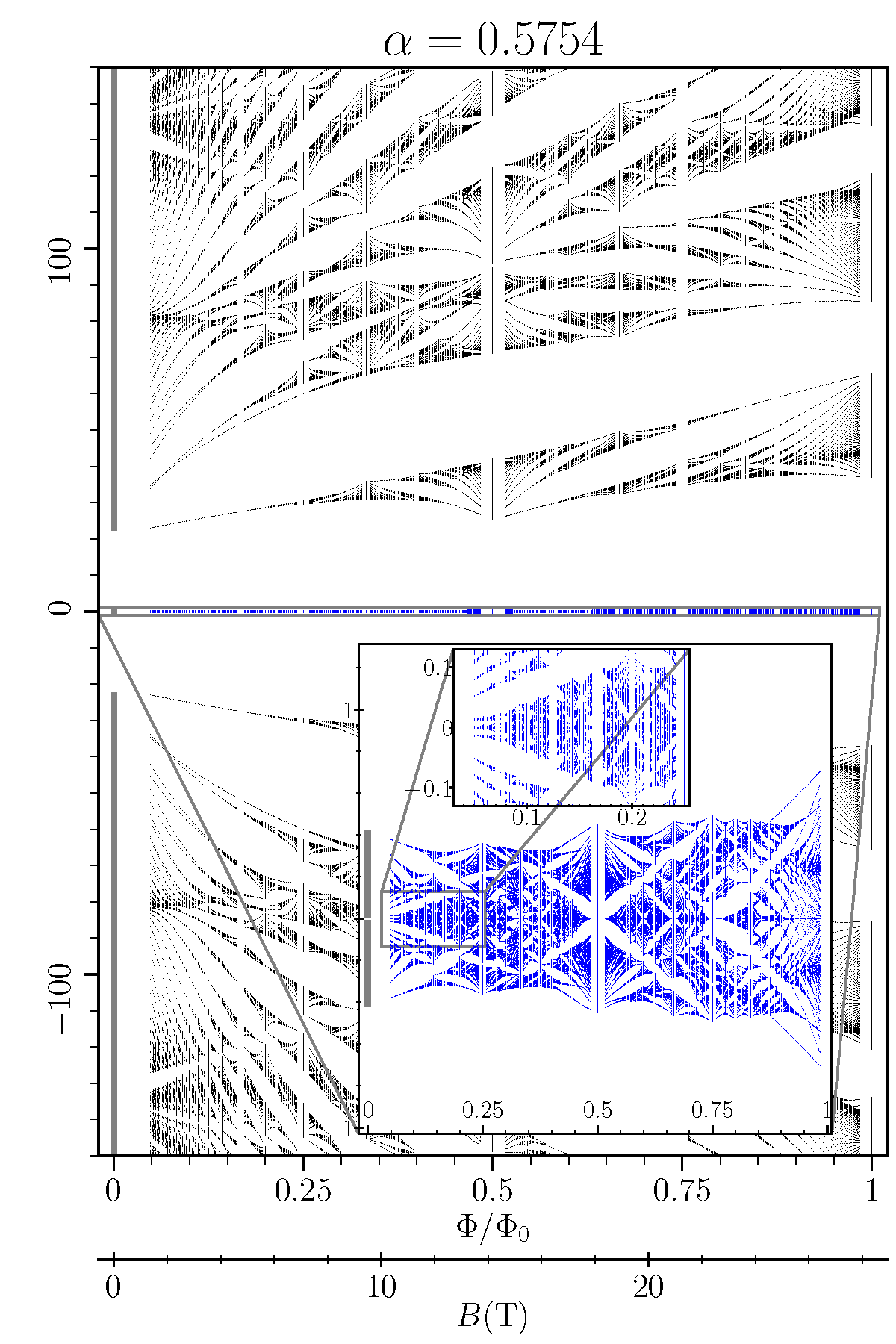}}
	\hspace{-0.1in}
	\subfigure[]{\label{fig:3500}\includegraphics[height=0.485\textwidth]{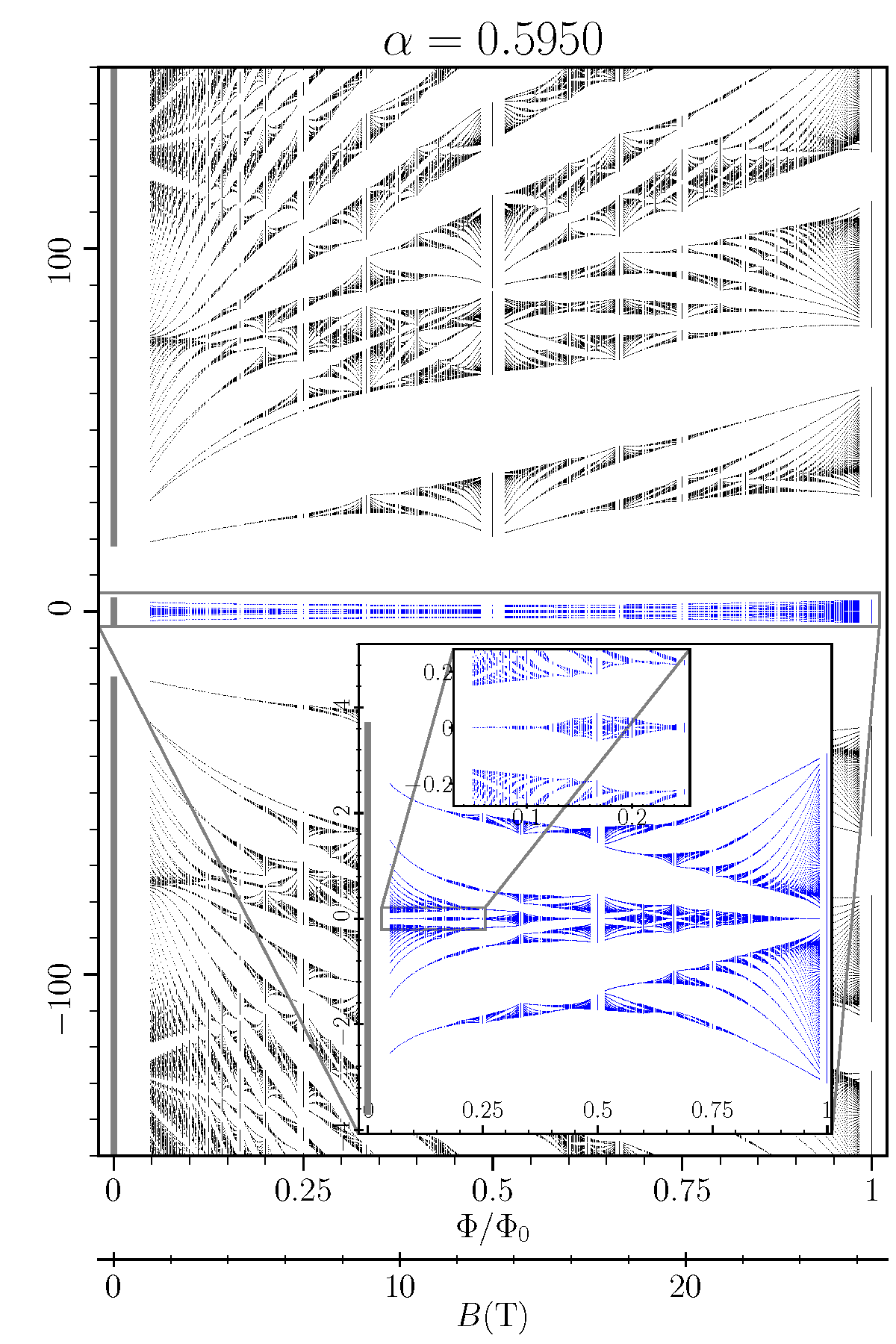}}
	\caption{\footnotesize Magnetic energy levels as a function of flux per moir\'e cell (the butterfly plot) for (a) $\alpha=0.35$ ($\theta\approx 1.82^\circ$), (b) $\alpha=0.5754$ ($\theta\approx 1.11^\circ$), and (c) $\alpha = 0.595$ ($\theta\approx 1.07^\circ$). The first $\alpha$ lies outside of and the other two lie within the magic range. The nonmagnetic bands in each case are also plotted on the far left with solid grey lines. The insets in (b) and (c) show magnified versions of the magnetic energy levels for the two active bands.}
	\label{fig:butterfly_plots}
\end{figure*}

We start with the following model Hamiltonian for zero magnetic field \cite{hejazi2019multiple}:
	\begin{equation}\label{eq:TBG_no_magnetic}
	\begin{aligned}
		H(\bm{x}) &= -i\, \left(\bm{\nabla} - i \tau^z \frac{\bm{q}_0}{2} + i \bm{q}_{\text{h}} \right) \cdot \left( \mathbf{R}_{\tau^z \theta/2} \, \bm{\sigma} \right) \\
		&+ \alpha \; \tau^+ \left[\eta \, \alpha_0(\bm{x})+\alpha_1(\bm{x}) \sigma^+ + \alpha_2(\bm{x}) \sigma^-\right] +\mathrm{h.c.},
	\end{aligned}
\end{equation}
where $\bm{x}$ and the Hamiltonian are made dimensionless by dividing by the moir\'e length scale $\frac{1}{k_\theta} = \frac{3\sqrt{3} a}{4\pi \, \theta}$ and the energy scale $\hbar v_{\text{F}} k_\theta$. The Pauli matrices $\tau^z=\pm 1, \sigma^z = \pm 1$ are used to denote top/bottom layer and A/B sublattice respectively. The constant vectors $\bm{q}_{\text{h}}=\left(\frac{\sqrt{3}}{2},0\right)$, and $\bm{q}_0 = \left( 0 , -1 \right)$ define the center of BZ and the tunneling term is defined using the functions $\alpha_n(\bm{x}) = \sum_{j=0}^2 e^{-i \bm{Q}_j . \bm{x}} \zeta^{nj}$, with $\zeta = e^{2\pi i/3}$. Also, $\bm{Q}_0 = 0$ and $\bm{Q}_1 = \sqrt{3} \left( -\frac12,\frac{\sqrt{3}}{2} \right)$ and $\bm{Q}_2 = \sqrt{3} \left( \frac12,\frac{\sqrt{3}}{2} \right)$ are the reciprocal moir\'e lattice vectors. The model has two parameters, one is $\alpha = \frac{w}{v_{\text{F}} k_\theta} \sim \frac{w}{\theta}$, which shows the combined effect of interlayer hopping and the twist angle and the other is $\eta$ which is responsible for incorporating the effect of corrugation \cite{koshino2018maximally}. This model concentrates on a single valley of graphene and a single spin, and so in order to take the complete physical system into account, four copies of $H$ should be introduced.   Upon neglecting the rotation of $\sigma$ matrices above, the effects of which are small for small $\theta$, one recovers a particle-hole symmetry as defined in Ref.~\onlinecite{hejazi2019multiple}. We will use this approximation unless otherwise stated.

Interestingly, there are two Dirac points (DPs) for all values of $\alpha$ at the $\text{K}$ and $\text{K}'$ points of the BZ. At the $\Gamma$ point, on the other hand, the top and bottom active bands reach their maximum and minimum respectively, except for a range of $\alpha$ around $\alpha_1 = 0.57544$ (for $\eta \approx 0.82$); at $\alpha_1$, the $\Gamma$ point becomes gapless with a quadratic band touching, and the bands show exceptional flatness. This transition happenning  at $\alpha_1$ is one of the series of topological transitions happening in the magic range ($\theta\approx 1.1^\circ$). A thorough study of the active bands' structure close to this transition and of the series of topological transitions happenning across the BZ is presented in Ref.~\onlinecite{hejazi2019multiple} and App.~\ref{app:nonmagnetic}.

We then incorporate the magnetic field in the same way as is done in Ref.~\onlinecite{bistritzer2011moire} (see App.~\ref{app:the_model} for a self-contained derivation), i.e.~we start by working in a basis of bare LLs of the two graphene sheets and take the effect of the tunneling term into account by finding its matrix elements in this basis. To solve the model, one is required to impose a commensurability condition for the magnetic flux and the moir\'e lattice, which validates the notion of a magnetic Brillouin zone (MBZ):
\begin{equation}
\frac{\Phi}{\Phi_0} = \frac{B \mathcal{A}}{\Phi_0} = \frac12 \frac{q}{p},
\end{equation}
where $\mathcal{A} = \frac{8\pi^2}{3\sqrt{3}k_\theta^2}$ is the moir\'e pattern unit cell area. \footnote{Note that the commensurability condition chosen here is different from the one in Ref. \onlinecite{bistritzer2011moire}, because of the difference in the convention for the moir\'e potential (see App.~\ref{app:the_model}).} This results in the relation $B(\mathrm{T}) \approx \frac{12 (\theta^\circ)^2}{p/q} \approx \frac{1}{\alpha^2}\frac{4.7}{p/q} $ between the magnetic field and the integers $p$ and $q$. Also, the Zeeman energy is neglected here (we will comment on the possible effect of Zeeman energy in Sec. \ref{sec:conclusion}). We only consider one of the valleys, and therefore the physical filling factor is 4 times that of the model one.  We will report the model filling factor unless otherwise stated.  
 
\section{Numerical solution and semiclassical analysis}\label{sec:num_sol_semi_classical}

In order to carry out numerical calculations, we need a cutoff for the LLs of the monolayer graphene sheets; we choose this cutoff by the criterion that the energies and the gaps found in the energy range of interest ($\left| E \right| \lessapprox 100 \mathrm{meV}$) remain approximately constant with further increase of the cutoff.   To achieve this condition, we find that the cutoff needs to be taken about ten times larger than the one introduced in Ref.~\onlinecite{bistritzer2011moire}. Generically, larger cutoffs are needed for smaller magnetic fields.

We work with $\eta \approx 0.82$ after Ref.~\onlinecite{koshino2018maximally}, and carry out the analysis for different values of $\alpha$. We start at larger angles than those in the magic range and discuss the results in depth, then using the same methods we specialize to the behavior in the magic range. The first $\alpha$ value we consider is $\alpha = 0.35$, which corresponds to $\theta = 1.8^\circ$. The energy spectrum as a function of the magnetic flux per unit cell (we will refer to such plots as butterfly plots) is shown in Fig. \ref{fig:butterfly_plots}. Note that the magnetic energies change continuously and form \textit{magnetic bands} as the magnetic Bloch momentum is varied within MBZ.
Each magnetic band when full corresponds to a density of electrons equal to $\frac{1}{\mathcal{A}}\frac{1}{2p}$, we call this quantity the weight\cite{wannier1978result} of that magnetic band (see App.~\ref{app:the_model}). 

One first observation from the above plot is as follows: for small enough magnetic fields, there are magnetic energy levels within the range of nonmagnetic active bands and nonmagnetic remote bands, but there is no magnetic energy level in the gap between them. This observation holds true for all $\alpha$ values where there is a gap in the nonmagnetic band structure, a range starting around $\alpha = 0.25$ and continuing up to around $\alpha = 0.65$. When this observation holds, by inspection in several cases, the total number of magnetic bands in the active range turns out to be $4p$; thus the corresponding total weight is equal to exactly two states per moir\'e supercell, which coincides with the density given by the active bands when full. We will be mostly studying the magnetic bands within this active energy range.

There are in principle gaps between each two adjacent magnetic bands, however, there are wider gaps between certain \emph{groups} of multiple bands. Such groups of bands can clearly be seen in the middle and at the edges of the active range of energy for small enough magnetic fields. By inspection, the groups of bands at the edges comprise $q$ bands each for a given $p/q$, while the groups of bands in the middle comprise $2q$ bands. Using the prescription above, it can be inferred that each of the former groups of bands carries a density of $\frac{1}{\mathcal{A}}\frac{q}{2p} = \frac{B}{\Phi_0}$ which is exactly the density of a full LL;
% at field strength $B$ 
 likewise, the latter groups carry 
% density of $\frac{1}{\mathcal{A}}\frac{2q}{2p} = 2\frac{B}{\Phi_0}$ which can be identified as
 the total density of two LLs. The wide gaps addressed here
%  between these groups
  should correspond to experimentally seen gaps; they persist over a finite range of $p/q$ and the weights above and below are continuous functions of the magnetic field. It is worthwhile here to also note that there are two groups of $q$ bands in the middle with energies close to zero, with a small gap between them, however this splitting grows for larger fields.

%\begin{figure*}[!t]
%	 \hspace{-0.1in}\centering
%	\subfigure[]{\label{fig:wannier_a}\includegraphics[height=0.26\textwidth]{wannier_alphaNo_3500_eta_817_thetadev_0_fit.png}}
%	\subfigure[]{\label{fig:wannier_b}\includegraphics[height=0.26\textwidth]{wannier_alphaNo_5754_eta_817_thetadev_0_fit.png}}
%	\subfigure[]{\label{fig:wannier_c}\includegraphics[height=0.26\textwidth]{wannier_alphaNo_5950_eta_817_thetadev_0_fit.png}} 
%	\subfigure{\includegraphics[height=0.26\textwidth]{wannier_alphaNo_0000_eta_817_thetadev_0_coloarbar.png}}
%	\caption{Wannier plots for (a) $\alpha=0.35$, (b) $\alpha=0.5754$, and (c) $\alpha = 0.595$. The vertical axes show the carrier density in the unit of one per moir\'e cell. The colors correspond to rescaled density of states $\rho/\rho_{\text{max}}$. The dark straight lines correspond to gaps and thus full LLs; the LL filling factors can be deduced using the slopes of these lines. A filling factor of $2$ for the zero energy LLs can be inferred by noting that at CNP there are dark lines with slopes $\pm 1$ in all of the plots, but no lines with slope $0$. Furthermore, the filling factors of $2$, $1$ and $3$ can be seen at CNP above the zero energy LL in the three plots respectively. \rd{Colorbar!}}
%	\label{fig:wannier}
%\end{figure*}

\begin{figure*}[!t]
	\centering
	\subfigure[]{\label{fig:wannier_a}\includegraphics[height=0.4\textwidth]{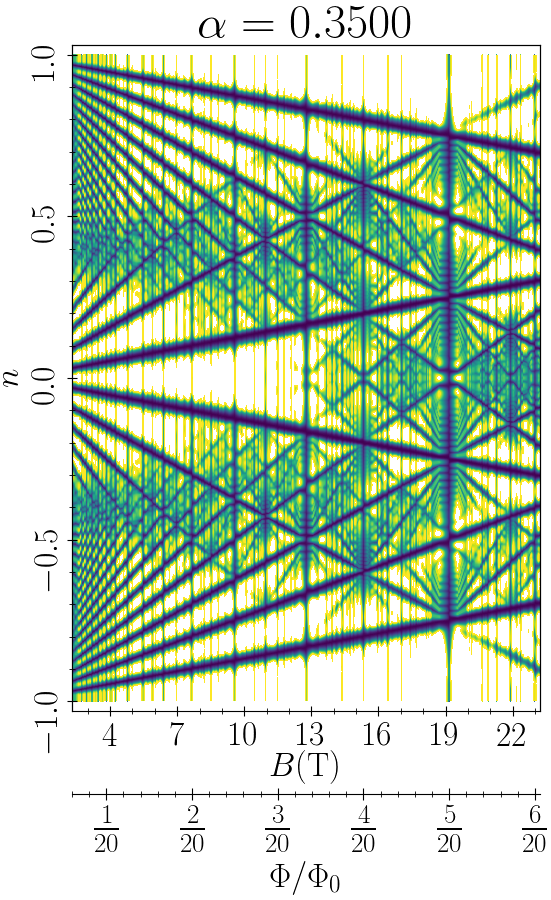}} \qquad
	\subfigure[]{\label{fig:wannier_b}\includegraphics[height=0.4\textwidth]{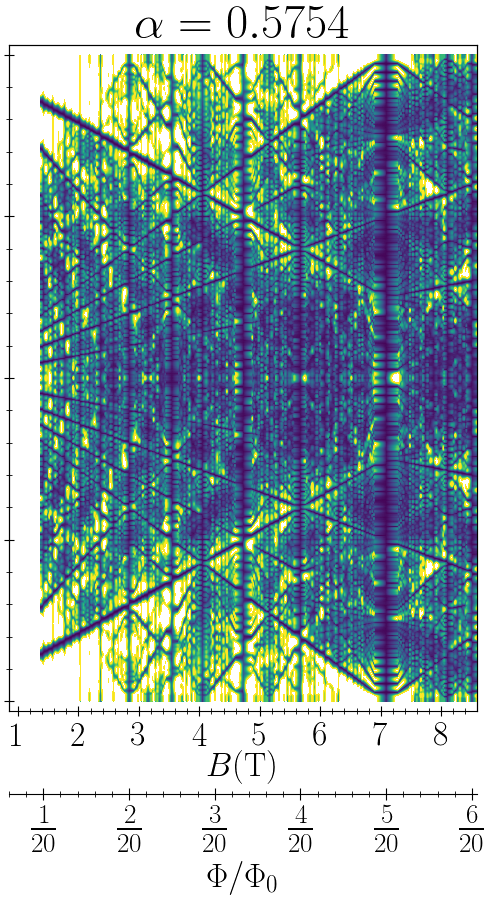}} \qquad
	\subfigure[]{\label{fig:wannier_c}\includegraphics[height=0.4\textwidth]{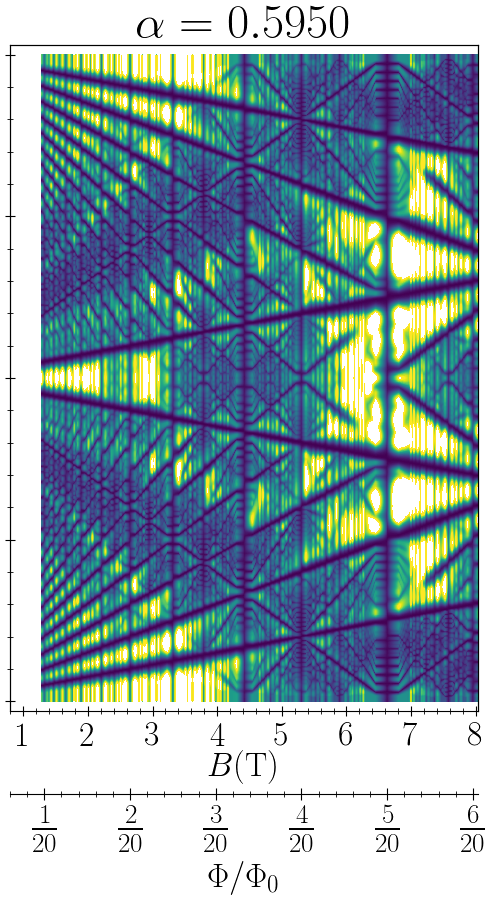}}  \qquad 
	\subfigure{\includegraphics[height=0.4\textwidth]{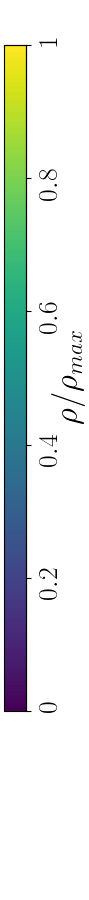}}
	\caption{Wannier plots for (a) $\alpha=0.35$, (b) $\alpha=0.5754$, and (c) $\alpha = 0.595$. The vertical axes show the carrier density in the unit of one per moir\'e cell. The colors correspond to rescaled density of states $\rho/\rho_{\text{max}}$. The dark straight lines correspond to gaps and thus full LLs; the LL filling factors can be deduced using the slopes of these lines. A filling factor of $2$ for the zero energy LLs can be inferred by noting that at CNP there are dark lines with slopes $\pm 1$ in all of the plots, but no lines with slope $0$. Furthermore, the filling factors of $2$, $1$ and $3$ can be seen at CNP above the zero energy LL in the three plots respectively. }
	\label{fig:wannier}
\end{figure*}

Another way to visualize the weights discussed above is using a Wannier plot\cite{wannier1978result}; a Wannier plot records the density of states $\rho$ as a function of carrier density $n$ and magnetic field $B$.  In order to calculate density of states we consider widening of each energy level found numerically by associating a Lorentzian density of states with it; Its width parameter $\gamma$  is chosen to ensure maximal resolution (see App.~\ref{app:wannier}). Such a plot for $\alpha = 0.35$ is presented in Fig.~\ref{fig:wannier_a}.
The energy gaps, corresponding to the minima in $\rho$, form straight lines; the LL filling factor can be inferred from the slope of these lines. Consistent with the above, filling factors of $2$ in the middle and $1$ at the edges can be deduced from this plot. Following all the above observations, the filling factor seen in Landau fan diagrams at larger angles \cite{cao2016superlattice}, can be understood by taking the above weights at CNP, resulting in the sequence $4 \times \left(\pm 1, \pm 3, \pm 5, \ldots \right)$ at the CNP (spin/valley degeneracy considered). Note that this coincides with the sequence found in Ref.~\onlinecite{cao2016superlattice}.

The formation of groups of bands and their weights for sufficiently small $B$ can be understood with a semiclassical analysis. In this method, in order to find the magnetic energy levels, one finds orbits of constant energy in $k$-space for a given band structure. Then one imposes the Bohr-Sommerfeld quantization condition\cite{alexandradinata2018semiclassical, chang1996berry}, which ultimately results in a quantization condition for the area in $k$-space enclosed by the orbits, denoted by $A_k$; the energy of such an orbit is obtained from the dispersion landscape. Concretely, quantized orbits formed around quadratic band edges satisfy the condition $A_k = 2\pi \frac{1}{\ell_{\text{B}}^2} \left( N + \frac12 \right) = \Omega_{\text{BZ}} \, \frac{\Phi_{\phantom{0}}}{\Phi_0} \left( N + \frac12 \right) $, and orbits enclosing a DP satisfy $A_k = 2\pi \frac{1}{\ell_{\text{B}}^2} \left( N + 1\right) = \Omega_{\text{BZ}} \, \frac{\Phi_{\phantom{0}}}{\Phi_0}  \left( N + 1\right)$, where $\ell_{\text{B}}$ is the magnetic length, and $ \Omega_{\text{BZ}} = \frac{3\sqrt{3}}{2} \, k_\theta^2$ is the total BZ area. The difference between the above two cases is due to the $\pi$ Berry phase accumulated when an orbit encloses a DP.

\begin{figure*}[t]
\centering
 \subfigure[]{\label{fig:dispersions_a} \includegraphics[width=.32\textwidth]{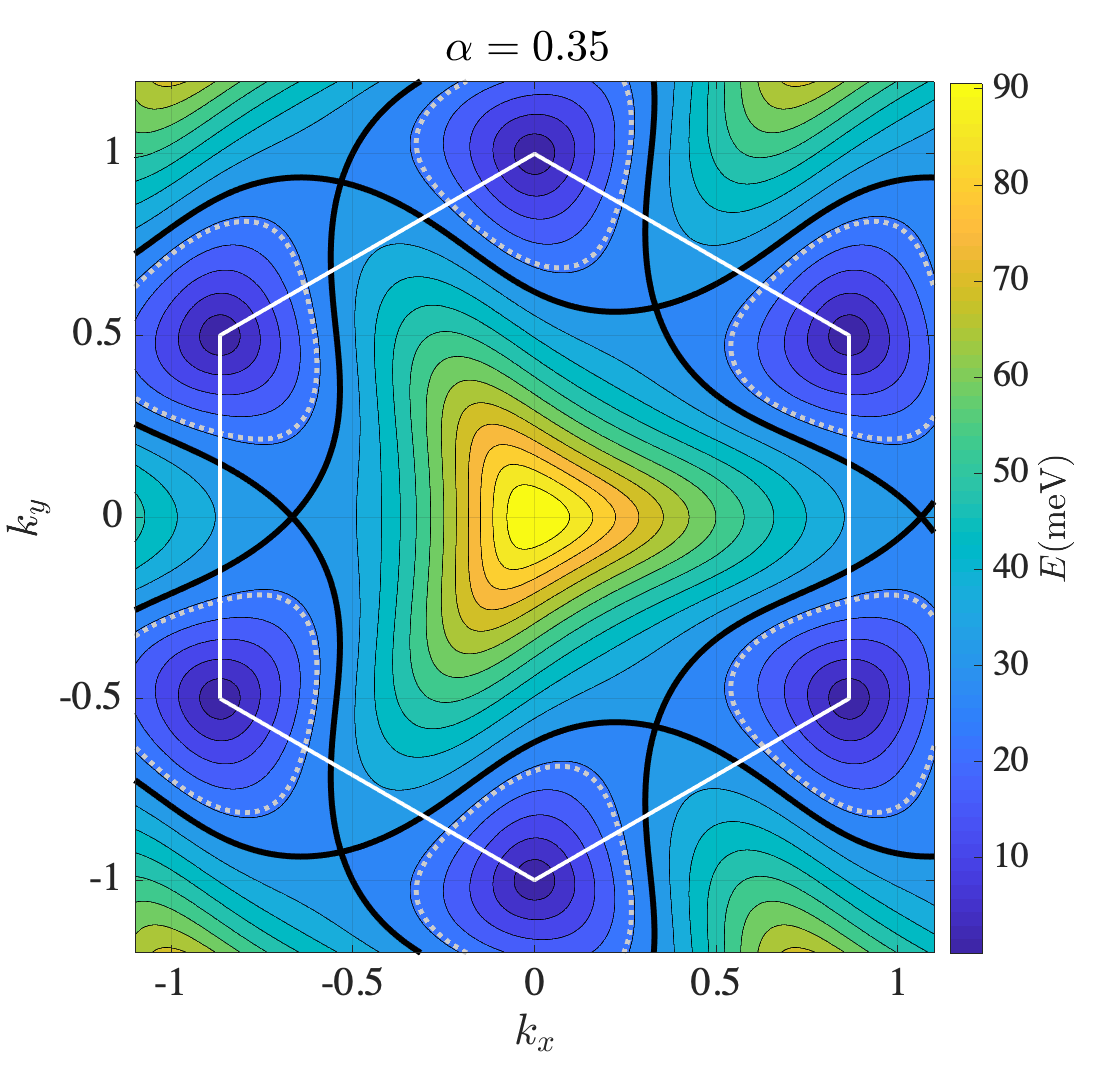}} 
 \subfigure[]{\label{fig:dispersions_b} \includegraphics[width=.32\textwidth]{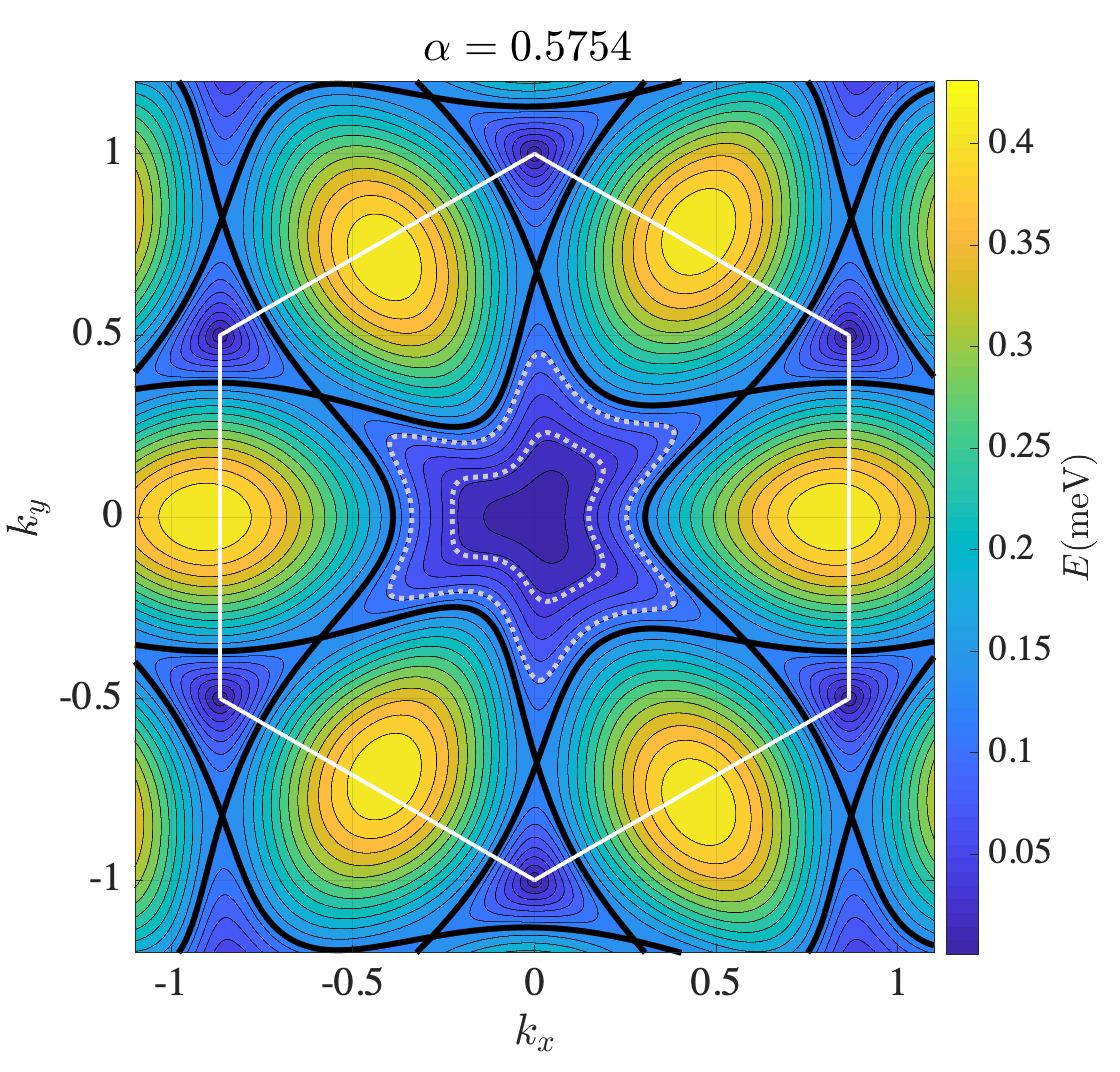}}
 \subfigure[]{\label{fig:dispersions_c}  \includegraphics[width=.32\textwidth]{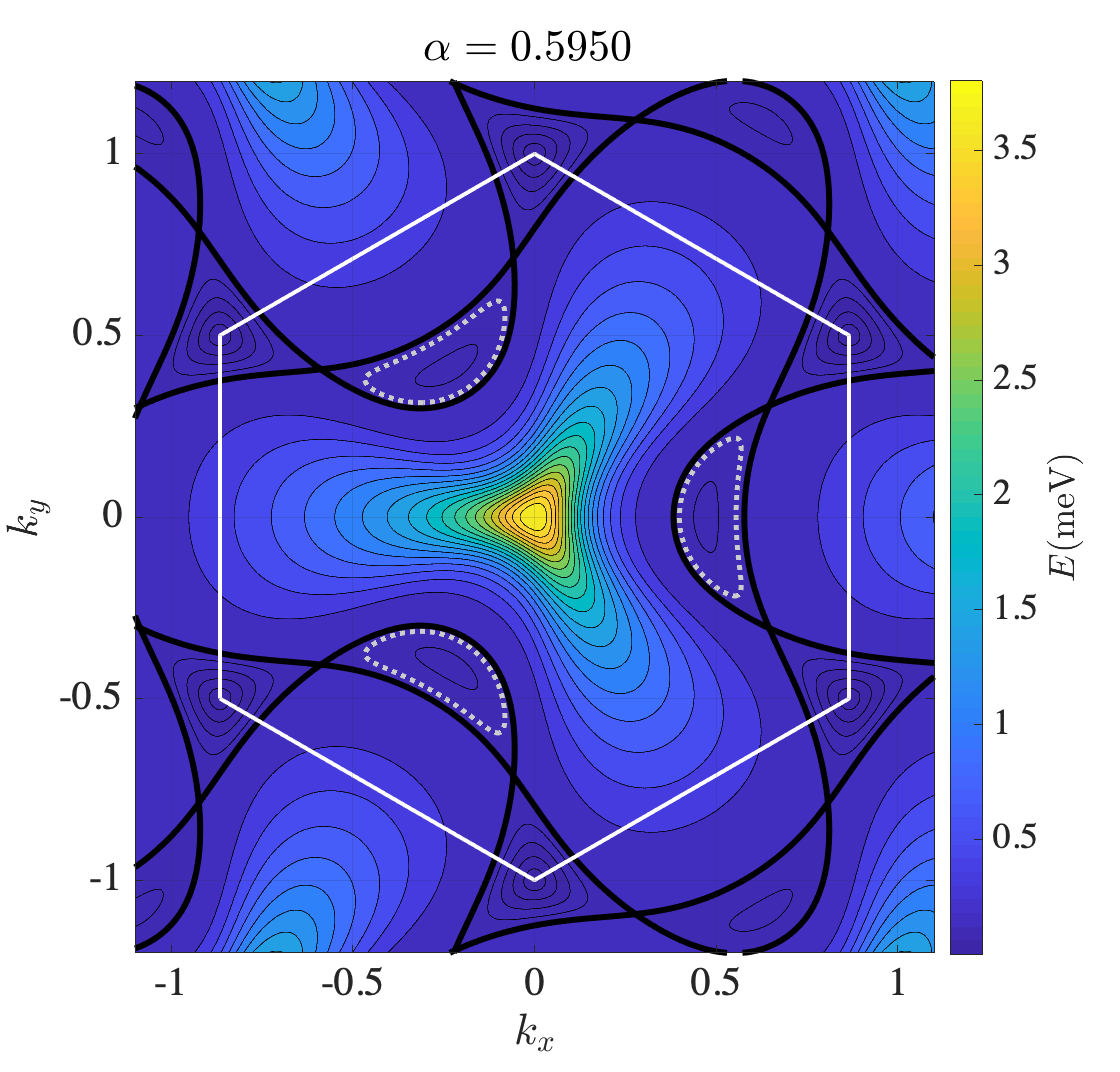}}
 \caption{ \label{fig:dispersions} The nonmagnetic band structure in the BZ for three values of $\alpha$. The BZ is shown as a white hexagon; there are two DPs located at its vertices $\text{K}$ and $\text{K}'$ points and the $\Gamma$ point is located at its center. It can be seen that in the magic range there are low energy regions in the BZ other than the vicinities of DPs (the two cases on the right). Some semiclassical orbits closest to CNP are shown in the three cases with dotted grey lines; the orbits are found at $\Phi/\Phi_0 = \frac{1}{10}$ for $\alpha = 0.35$, and $\alpha = 0.5754$,  at $\Phi/\Phi_0 = \frac{1}{25}$ for $\alpha = 0.595$.} 
\end{figure*}

A plot of the nonmagnetic dispersion at $\alpha = 0.35$  in the top active band is presented in Fig.~\ref{fig:dispersions_a}; 
the two classes of groups of magnetic bands that are formed in the active range i.e. $q$-band groups at the edge and $2q$-band groups in the middle can be associated with the semiclassical orbits formed in this dispersion surface; each group in the former class can be identified as an orbit around the $\Gamma$ point and each group in the latter class as the collection of two orbits each around one of the moir\'e Dirac cones (points $\text{K}$ and $\text{K}'$). There is very good quantitative agreement between the energies found this way and the energies found in the butterfly plots for small enough fields (see App.~\ref{app:semiclassics}).

We then specialize to the discussion of the magnetic bands in the magic range, i.e.~where the active energy range becomes very narrow. The above scenario for the weights in the middle and the edges of the active range remains valid upon increasing $\alpha$ until $\alpha$ gets close to $\alpha_1$ where the nonmagnetic active bands touch at the $\Gamma$ point and show exceptional flatness. Noting that $\alpha_1 = 0.57544$, we present results at the close value of $\alpha = 0.5754$, but as discussed later there is in fact a range of $\alpha$ where this applies. A butterfly plot for $\alpha = 0.5754$ ($\theta \approx 1.1^\circ$) is presented in Fig.~\ref{fig:butterfly_plots}. Still there are $2p$ bands in total in the active range, corresponding to two electrons per moir\'e cell.

As before, groups of bands with wide gaps between them can be seen at the edges and in the middle of the active range. This time however, their group weights are different; each of the groups formed in the middle consists of $q$ bands and each of the ones at the edges contains $3q$ bands. Since a $q$-band group corresponds to the weight of a LL, each of the middle groups carry the density of one LL and each of the edge groups carry the density of three LLs. The two zero energy LLs still have the same behavior. These weights are also exhibited in the Wannier plot shown in Fig.~\ref{fig:wannier}; as a result, one expects a sequence $4 \times \left( \pm 1,\pm 2, \pm 3, \ldots \right)$ at CNP taking spin and valley degeneracy into account.  

A dispersion plot of the top active band is shown in Fig.~\ref{fig:dispersions_b} for $\alpha = 0.5754$; note that there is a quadratic low energy dispersion at the $\Gamma$ point. Remarkably, a semiclassical analysis shows that each of the middle groups of bands corresponds to a semiclassical orbit around the $\Gamma$ point.
Also, semiclassical analysis relates the groups with weights of three LLs at the edges to the orbits around the three inequivalent points $\text{M}_1,\text{M}_2,\text{M}_3$ (which are  $C_3$ related), where the highest (lowest) energy of the top (bottom) nonmagnetic band is reached with a quadratic dispersion (see App.~\ref{app:semiclassics} for details).
This set of LL filling factors can be seen in the $\alpha$ range $0.575$--$0.585$, where there can be several DPs in the nonmagnetic active bands; we also observe several level crossings as the magnetic field is varied for small fields, but the sequence is unchanged away from the band crossing points.

Finally, for slightly larger values of $\alpha$, yet another different pattern for LL degeneracies emerges, as can be seen in Fig.~\ref{fig:butterfly_plots} which shows results for $\alpha = 0.595$ ($\theta \approx 1.1^\circ$): a group of bands with a weight equal to three LLs appears in the middle of the active range, and groups of weight equal to a single LL appear at the edge. A semiclassical study (Fig.~\ref{fig:dispersions_c}) shows that the group of three LLs in the middle can be identified as the collection of the orbits that take place in the three low energy quadratic dispersions centered on high symmetry lines. On the other hand, the edge groups can be identified as the orbits enclosing the $\Gamma$ point. The former results in the sequence $4 \times \left(\pm 1, \pm 4, \ldots \right)$ at CNP. This set of LL filling factors can be observed starting around $\alpha=0.59$ up to $ \alpha \approx 0.64$, where the nonmagnetic gap is closed and even beyond that.

It appears from the above results, and in particular the semiclassical analysis, that in the magic range, the low energy magnetic levels around CNP are not related to the dispersion around the two moir\'e DPs; in fact, one can get further insight using semiclassical analysis: one can identify the only contour in BZ that intersects itself as the saddle contour, which can play the role of separating different classes of orbits based on the orbit centers (Fig. \ref{fig:dispersions}). In all the above cases, the saddle contours enclose the DPs and therefore limit the total area available to the orbits forming around $\text{K}$ and $\text{K}'$; 
it is indeed the case at $\alpha = 0.5754$ and $\alpha = 0.595$ that the total area available around each DP is sufficient for an orbit to form only for very small magnetic fields, while orbits within other quadratic dispersion areas of the BZ begin to form at much larger fields. Concretely, the first orbits around the DPs only form at $\frac{\Phi}{\Phi_0} \approx \frac{1}{25}$ ($B \approx 1 \text{T}$) in the first case and at $\frac{\Phi}{\Phi_0} \approx \frac{1}{30}$ in the second case.

%\begin{figure*}[t!]
%	\centering
%	\begin{subfigure}[t]{0.325\textwidth}
%		\centering
%		\includegraphics[width=\textwidth]{dispersion_3500.png}
%		\caption{\label{fig:dispersions_a}  }
%	\end{subfigure}%
%~%
%	\begin{subfigure}[t]{0.325\textwidth}
%		\centering
%		\includegraphics[width=\textwidth]{dispersion_5754.png}
%		\caption{\label{fig:dispersions_b} }
%	\end{subfigure}
%~%
%\begin{subfigure}[t]{0.325\textwidth}
%	\centering
%	\includegraphics[width=\textwidth]{dispersion_5950.png}
%	\caption{\label{fig:dispersions_c} }
%\end{subfigure}
%	\caption{\label{fig:dispersions} The nonmagnetic band structure in the BZ for three values of $\alpha$. It can be seen that there are low energy regions other than the vicinity of DPs in the BZ for the two cases on the right. Semiclassical orbits closest to CNP are shown in the three cases; the orbits are found at $\Phi/\Phi_0 = \frac{1}{10}$ for $\alpha = 0.35$, and $\alpha = 0.5754$,  at $\Phi/\Phi_0 = \frac{1}{25}$ for $\alpha = 0.595$. }
%\end{figure*}

\section{Discussion and Conclusion}\label{sec:conclusion}
We have considered the non-interacting continuum model of twisted bilayer graphene at twist angles above 
%($\approx 2^\circ$) It is not the case that we have considered the single value of $2^\circ$ only and so it is better not to metnion this
 and within the magic range ($\approx 1^\circ$) when subject to a perpendicular magnetic field. The magnetic energy levels are found by imposing magnetic commensurabilty with the moir\'e superlattice, resulting in butterfly plots. By careful inspection we have derived three different sequences for the LL filling fractions at CNP, i.e.~$\nu = \pm 4, \pm 12, \pm 20, \ldots$, for angles larger than the magic angle, and $\nu = \pm 4, \pm 8, \pm 12, \ldots$ and $\nu = \pm 4, \pm 16, \ldots$ within the magic range with spin and valley degeneracies considered; we have presented these results concretely at the three $\alpha$ values of $0.35$, $0.5754,$ and $0.595$ respectively.  We note that the second of these sequences seems to correspond to that which is observed experimentally by several groups.  However, in the present model this sequence occurs only for a narrow range of $\alpha$ parameter.  It is possible that the range in which this behavior occurs is enhanced by interaction effects, not included here, for example self-energy corrections.  It is also possible that the observed sequence has an entirely different origin.  We do not resolve this here but believe our results provide useful input to the issue.

We explained the above results by a semiclassical analysis of the energies at small magnetic fields.  We found that in the magic angle range there is not enough area for even the first semiclassical orbit to form around the moir\'e DPs, unless the magnetic field is very small ($\lessapprox 1\text{T}$). This shows that in the magic range other areas of the BZ are responsible for the low energy LLs around CNP; as discussed above, the $\Gamma$ point for $\alpha = 0.5754$ and three local minima appearing on high symmetry lines for $\alpha = 0.595$ play this role.

We have neglected the effect of Zeeman energy in this work; this approximation is indeed justified by an experimental result in Ref.~\onlinecite{yankowitz2019tuning} where it is argued, by a comparison between observed landau fan diagrams with a perpendicular field and a titled field, that the main sequence seen in Landau fan diagrams is not caused by spin splitting. However, taking the Zeeman effect into account will be also interesting.

Although we have been mostly concerned with relatively small magnetic fields in this work, the formalism also works for large fields. In particular, we have studied how the transition from the previously discussed small $B$ regime to large $B$ regime happens when there is a gap between active and remote nonmagnetic bands (see App.~\ref{app:duality_adiabaticity}). 
%in the former case, one recovers the semiclassical orbits in the nonmagnetic active bands, while for the latter, one expects to recover bare LLs of the monolayer graphene sheets, each widened due to the effect of moir\'e lattice. 
Deep in the small field limit, as discussed above we see a total density equal to two electrons per moir\'e cell confined within the active range of energy; the two gaps above and below this range evolve continuously as the magnetic flux is increased. 
In the large field limit on the other hand, one expects to recover the bare LLs of the monolayer graphene sheets, each widened due to the effect of the moir\'e lattice. 
The interpolation between the above two limits happens at intermediate field range; at a generic $\eta$ (we have been using $\eta = 0.82$ as an example) one can see two different behaviors as $\alpha$ is varied:
 either i) the zeroth LLs of the two moir\'e DPs (small $B$) are adiabatically connected to the bare zeroth LLs of monolayer sheets (large $B$), which happens for smaller gaps, or ii) the total weight of two nonmagnetic bands (active range) at small $B$ is adiabatically continued into the large $B$ limit also, which occurs when the gap becomes larger (see Figs.~\ref{fig:full} and \ref{fig:adiabaticity} in App.~\ref{app:duality_adiabaticity} for these cases). This observation is consistent with the results presented in Ref.~\onlinecite{lian2018landau}, especially those where there is a gap\footnote{
	The result was reported differently in an earlier version of Ref.~\onlinecite{lian2018landau}: it was stated that the small field gap between the magnetic bands confined within the active range and higher bands is always closed with a sharp transition around $\Phi / \Phi_0 = 1$, for all $\alpha$ values. This was attributed to the ``fragile topology'' of the active bands in TBG.
}.

As stated above, one expects to recover bare LLs of monolayer graphene sheets in the infinite magnetic limit; the two bare zeroth LLs  are infinitely far apart in energy from other LLs, and so in this limit a low energy description of the model can be obtained through projection onto these two LLs.
Remarkably, by looking at the detailed structure of the projected Hamiltonian, we find a duality between the description of these zeroth LLs at infinite magnetic flux limit, and a tight binding honeycomb model (i.e. the honeycomb Hofstadter butterfly\cite{rammal1985landau}) at small magnetic flux limit (see App.~\ref{app:infinite_B} for detail), with the layer index for the two zeroth LLs $s=1,2$ in the former theory playing the rule of honeycomb sublattice index in the latter. Sharing the same band structure and density of states, the latter theory which has been studied extensively can shed some light on the expected properties of former (see Appendices \ref{app:infinite_B} and \ref{app:chern} for details).

It is worthwhile to discuss also the particle-hole symmetry we have considered here; restoring the sublattice pseudospin rotation in both the magnetic and nonmagnetic Hamiltonians will result in breaking of the particle-hole symmetry of both spectra. For the nonmagnetic model, it is worth mentioning that restoring the sublattice pseudospin rotation precludes a Lifshitz transition from happening at or in the vicinity of $\mathrm{K}$ and $\mathrm{K}'$ and thus the Dirac velocity at these points does not vanish at any $\alpha$ value for nonzero $\eta$ (see App.~\ref{app:nonmagnetic} and Fig.~\ref{nonB} therein for detailed analysis). While this observation invalidates a conventional definition of the magic angle as a single angle where the Dirac velocity at $\text{K}$ and $\text{K}'$ vanishes, it is still legitimate to talk about a \emph{magic range} of angles in which the two active bands show considerable flatness. Additionally, in the magnetic model, we have checked that (see App.~\ref{app:ph_broken} for details) the relatively small field results and in particular the filling factors do not change except when one is close to $\alpha_1$, where some level crossings can occur at small magnetic fields in the middle of the active range (see Fig.~\ref{fig:5754_thetadev_1} in App.~\ref{app:ph_broken}). These level crossings can be understood by noting that there are $\alpha$ values at which both of the non-magnetic active bands have a minimum (or a maximum) at the $\Gamma$ point; as a result, an orbit forming around the $\Gamma$ point in the top layer can have an energy smaller than that of the zeroth LLs of the DPs for small magnetic fields (see Fig.~\ref{fig:nonmagnetic_broken_ph} in App.~\ref{app:ph_broken}).

It is natural to expect that the LL degeneracy of $3$ (not taking spin/valley degeneracy into account) that is seen at the edge of the active range at $\alpha = 0.5754$ and in particular at CNP at $\alpha = 0.595$ can be lifted when other effects are taken into account to make the study more realistic. The following three are the most obvious effects to consider: i) the effect of symmetry breaking terms at the level of noninteracting physics which can be induced by the effects of the environment, such as the hBN substrate; ii) the effect of disorder, which is not taken into account here and can have very nontrivial impact on Dirac dispersions\cite{nomura2007topological,bardarson2007one}; iii) and finally the effect of electron-phonon and electron-electron interactions which are neglected here. It would be an interesting further step to explore how taking these effects into account can affect the results presented here.

\begin{acknowledgements}  
	We thank A. Young, G. Polshyn, B. Lian, A. Macdonald, C. Murthy and X. Wu for fruitful discussions we have had with them.
We acknowledge support from the Center for Scientific Computing from the CNSI, MRL: an NSF MRSEC (DMR-1720256) and NSF CNS-1725797. The research of the authors was supported by the National Science Foundation grant DMR1818533.

\end{acknowledgements}

\bibliography{graphene}

\onecolumngrid
\appendix
%\section*{\Large supplementary information}

\section{The model}\label{app:the_model}
The Hamiltonian consists of two terms:
\begin{equation}
	H = H_{\text{LL}} + H_{\text{tunneling}}.
\end{equation} 
We have made everything dimensionful here to keep track of the new parameters in terms of the magnetic field, however we will ultimately work with the dimensionless Hamiltonian $H/(\hbar v_{\text{F}} k_\theta)$, as in the nonmagnetic case.

\begin{itemize}
	\item  $H_{\text{LL}}$ is the Hamiltonian corresponding to the bare LLs of each of the graphene sheets and can be written in the following form:
\begin{equation}
	\begin{aligned}
		H_{\text{LL}} = \hat{P}_+ \, h(-\theta / 2) + \hat{P}_- \, h(\theta / 2).
	\end{aligned}
\end{equation}
where $\hat{P}_{\pm} = \frac{1 \pm \tau^z}{2}$. The single layer Hamiltonian $h(\theta/2)$ is also defined as:
\begin{equation}
	h(\theta/2) = \hbar v_{\text{F}}\, \left[ - i \nabla + e \bm{A} + \mathrm{sgn}(\theta) \frac{\bm{q}_0}{2} + \bm{q}_h \right] \cdot \left( \mathbf{R}_{\theta/2} \, \bm{\sigma}   \right).
\end{equation}
$\mathbf{R}$ is a rotation matrix:
$$\mathbf{R}_{\theta/2} = \begin{pmatrix}
	\cos \theta/2 & -\sin \theta/2 \\
	\sin \theta/2 & \cos \theta/2
\end{pmatrix}.$$

The single layer Hamiltonian can finally be written as:
%\begin{equation}
%h(\theta / 2) = \hbar v_{\text{F}} \,  \left[ \sigma^+ e^{i \theta /2} \left(\sqrt{\frac{2eB}{\hbar}} O -i \text{sgn}(\theta)\frac{q_{0,y}}{2}+ q_{h,x}\right) + h.c.\right],
%\end{equation}
\begin{equation}
h(\theta / 2) = \hbar v_{\text{F}} \,  \left[ \sigma^+ e^{i \theta /2} \left(\sqrt{\frac{2eB}{\hbar}} O + i \,  \text{sgn}(\theta)\frac{\left|\bm{q}_{0}\right|}{2}\right) + \text{h.c.}\right],
\end{equation}
the operators $O$ and $O^\dagger$ are defined as follows using the Landau gauge $\bm{A} = B \left( -y , 0 \right)$ :
\begin{equation}
	\sqrt{\frac{2eB}{\hbar}} O = - \partial_y + k_x + \left| \bm{q}_h \right| - \frac{eBy}{\hbar}, \qquad \sqrt{\frac{2eB}{\hbar}} O^\dagger = \partial_y + k_x +  \left| \bm{q}_h \right| - \frac{eBy}{\hbar}.
\end{equation}
They are raising and lowering operators of LL index:
\begin{equation}\label{eq:commutation_O_Od}
	\left[ O , O^\dagger \right] = 1.
\end{equation}
The wave functions are extended in the $x$ direction and harmonic-oscillator-like (localized) in the $y$ direction.

	\item $H_{\text{tunneling}}$ can be found by computing the matrix elements of the tunneling terms in the LLs found above. 
	\begin{itemize}
		\item	The commensurability condition is taken as follows ($\ell_{\text{B}} = \sqrt{\frac{\hbar}{e B}}$):
\begin{equation}
	\frac32 k_\theta \Delta = \frac{3\sqrt{3}}{4} k_\theta^2 \ell_{\text{B}}^2 = 2 \pi p/q,
\end{equation}
where $\Delta$, the change in the guiding center induced by the tunneling term is given by:
$$\Delta = \sqrt{3} \, k_\theta \ell_{\text{B}}^2 / 2.$$
	Note furthermore that since we are working with a different but equivalent form of the tunneling, our commensurability condition is different. In terms of the moir\'e pattern unit cell area, the above condition can be written as:
	\begin{equation}
\frac{B \mathcal{A}}{\Phi_0} = \frac12 \frac{q}{p}.
\end{equation}

	\item We will work with a basis of LL's as follows: $|\tau,n,\sigma,y_c \rangle$, where $\tau$ shows the layer, $n$ shows LL index, $\sigma$ shows sublattice and $y_c$ is the guiding center coordinate. The guiding centers in a tunneling process can only change with the values $\pm \Delta$.
Thus one can write $y_c$ as $y_c = y_0 + \left( mq + j \right) \Delta$, with 
$$0<y_0 = k_1 \ell_{\text{B}}^2 < \Delta, \qquad 0<j<q-1,$$
and $j = j + q$. The parameter $k_1$ defines the $x$ component of the magnetic Bloch  momentum. Then we can do a Fourier transform on the parameter $m$ and work with the new basis:
\begin{equation}
	|\tau,n,\sigma,y_0,j,k_2 \rangle = \frac{1}{\sqrt{N}} \sum_m e^{i k_2 (mq+j)\Delta} |\tau,n,\sigma,y_0+(mq+j)\Delta\rangle.
\end{equation}
The parameter $k_2$ defines the $y$ component of the magnetic Bloch momentum.
\item The tunneling term can be decomposed into three terms according to the different spatial dependences:
\begin{equation}
	H_{\text{tunneling}} = \left(\mathcal{T}_0 + \mathcal{T}_1 + \mathcal{T}_2  \right)+ \text{h.c.}.
\end{equation}
Each term in the above form $\mathcal{T}_n$ is given by 
\begin{equation}
\frac{1}{\hbar v k_\theta}\mathcal{T}_n = \tau^{+} \alpha \, e^{-i \,\bm{Q}_n \cdot \bm{x}} \, T_n ,
\end{equation}
where $\bm{Q}_0 = 0$, $\bm{Q}_1 = \sqrt{3} \, k_\theta \left( -\frac12,\frac{\sqrt{3}}{2} \right)$ and $\bm{Q}_2 = \sqrt{3} \, k_\theta \left( \frac12,\frac{\sqrt{3}}{2} \right)$ and the $2\times2$ matrices $T_n$ are given by:
%\begin{gather}
%T_0 = \eta \, \sigma^0 + \sigma^x, \quad T_1 = (\eta \zeta^*)  \, \sigma^0 + \sigma^+ + \zeta \, \sigma^-, \quad T_1 = (\eta \zeta)  \, \sigma^0 + \sigma^+ + \zeta^* \, \sigma^-,\\
%\left( \zeta = e^{2\pi i /3}  \right),\nonumber
%\end{gather}
\begin{gather}
T_0 = \eta \, \sigma^0 + \sigma^x, \quad T_1 = \eta  \, \sigma^0 +  \zeta\, \sigma^+ + \zeta^* \, \sigma^-, \quad T_1 = \eta  \, \sigma^0 + \zeta^* \, \sigma^+ + \zeta \, \sigma^-,\\
\left( \zeta = e^{2\pi i /3}  \right),\nonumber
\end{gather}

\item	Now, each of the tunneling terms in the dimensionless form can be written as follows:
\begin{equation}\label{eq:tunnelingterms}
	\begin{aligned}
\frac{1}{\hbar v_{\text{F}} k_\theta}\langle 1,n',\sigma',y_0',j',k_2'| \,& \mathcal{T}_0 \,|2,n,\sigma,y_0,j,k_2\rangle = (\alpha) \ \delta_{y_0,y_0'}\,\delta_{jj'} \,\delta_{k_2,k_2'} \, \delta_{n',n} \, \langle 1,\sigma'|\, T_0 \,|2,\sigma\rangle,\\
\frac{1}{\hbar v_{\text{F}} k_\theta}\langle 1,n',\sigma',y_0',j',k_2'| \,& \mathcal{T}_1 \,|2,n,\sigma,y_0,j,k_2\rangle = (\alpha) \ \delta_{y_0,y_0'}\, \delta_{(j+1)j'} \,\delta_{k_2,k'_2}\\
& \times \ \langle 1,\sigma'|\, T_1 \,|2,\sigma\rangle \ F_{n'n}(\widetilde{\bm{Q}}_1 \, \ell_{\text{B}}/\sqrt{2}) \ e^{-\frac{3}{2} i k_\theta y_0}\, e^{-i \, k_2 \Delta } \  e^{ - i \frac{2\pi p}{q} \left(j+\frac{1}{2}\right)}\, ,\\
\frac{1}{\hbar v_{\text{F}} k_\theta}\langle 1,n',\sigma',y_0',j',k_2'| \,& \mathcal{T}_2 \,|2,n,\sigma,y_0,j,k_2\rangle = (\alpha) \ \delta_{y_0,y_0'}\, \delta_{(j-1)j'} \,\delta_{k_2,k'_2}\\
& \times\ \langle 1,\sigma'|\, T_2 \,|2,\sigma\rangle \ F_{n'n}(\widetilde{\bm{Q}}_2 \, \ell_{\text{B}}/\sqrt{2}) \ e^{-\frac{3}{2}i k_\theta y_0} \, e^{i \, k_2 \Delta } \ e^{ -i \frac{2\pi p}{q} \left(j-\frac{1}{2}\right)}\, .
\end{aligned}
\end{equation}
Where in the above equations $\widetilde{\bm{Q}}_j = Q_{j,x} + i \, Q_{j,y}$.

\item The function $F$ reads:
\begin{equation}
	F_{n'n}(z) = \begin{cases} 
      (-z^*)^{n'-n} \sqrt{\frac{n!}{n'!}} \, L_n^{n'-n}(z z^*) \, e^{-z z^*/2} & n'\geq n \\
      (z)^{n-n'} \sqrt{\frac{n'!}{n!}} \, L_{n'}^{n-n'}(z z^*) \, e^{-z z^*/2} & n'< n 
   \end{cases}\ .
\end{equation}
$L_a^b$ is the generalized Laguerre function.

	\item The magnetic BZ for the magnetic momentum $\bm{k}=(k_1,k_2)$ is a region given by:
	\begin{equation}\label{eq:magneticBZ}
0 < k_1 k_\theta \ell_{\text{B}}^2 = k_\theta y_0 < \frac{4\pi}{3} \frac{p}{q}, \qquad \quad 0 < k_2 \Delta<   \frac{2 \pi}{q}.
\end{equation}

\item Using the commensurability condition the dimensionless single layer term can also be written as:
\begin{equation}
\frac{1}{\hbar v_{\text{F}} k_\theta} h(\theta/2) =  \sigma^+ e^{i \theta /2} \left(
\sqrt{\frac{3\sqrt{3}}{4\pi} \, \frac{q}{p}} \ O +\frac{i \,\text{sgn}(\theta)}{2}\right)
+ \text{h.c}.
\end{equation}

\end{itemize}

\end{itemize}

\section{Small and large magnetic field correspondence in the butterfly plots}
\label{app:duality_adiabaticity}

\begin{figure}[!thb]
\centering
\subfigure[]{\label{fig:full250}\includegraphics[width=0.49\textwidth]{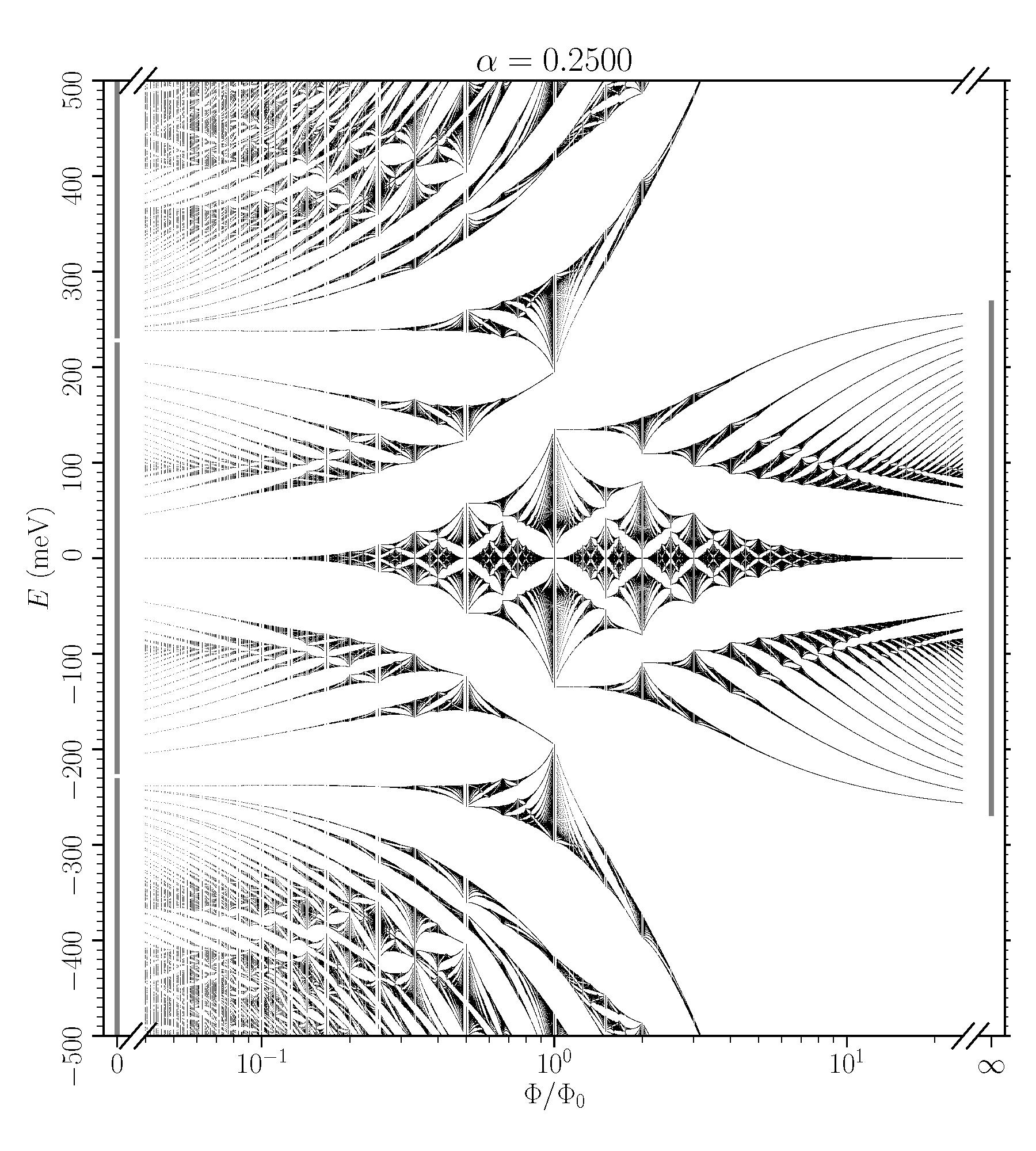}}
\subfigure[]{\label{fig:full350}\includegraphics[width=0.49\textwidth]{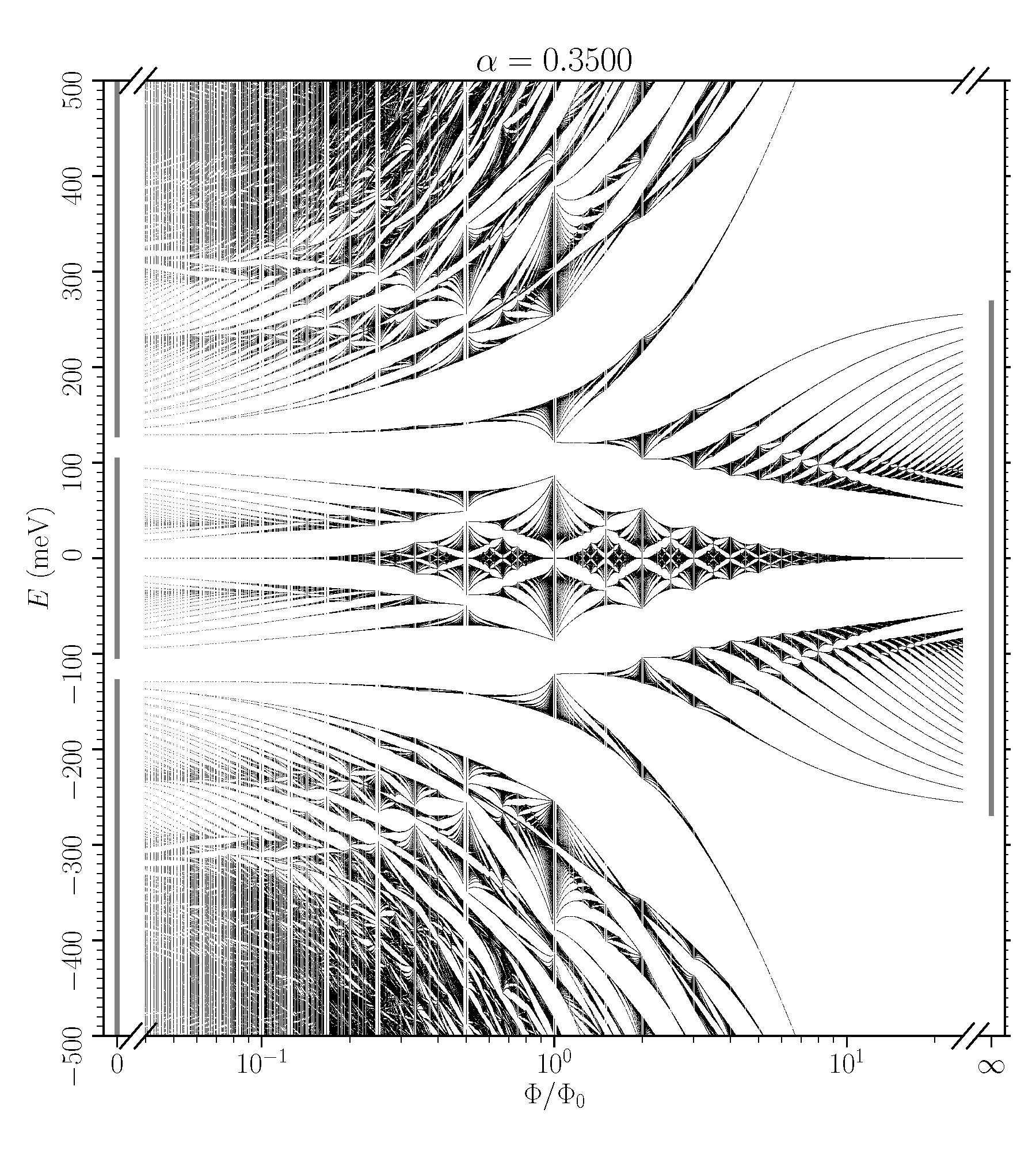}}
\caption{Butterfly plots in the wide range of $\frac{1}{25}\leq \Phi/\Phi_0\leq 25$ for (a) $\alpha = 0.25$ ($\theta = 2.5^\circ$) and (b) $\alpha = 0.35$ ($\theta=1.8^\circ$). The horizontal axis uses log scale to reflect the duality between $\Phi/\Phi_0$ and $\Phi_0/\Phi$. The bands for the nonmagentic case and the infinite magnetic field case (see App.~\ref{app:infinite_B}) are also plotted with solid grey lines, on the far left and far right, respectively.}\label{fig:full}
\end{figure}

\begin{figure}[!thb]
\centering
\subfigure[]{\label{fig:250analysis}\includegraphics[width=0.73\textwidth]{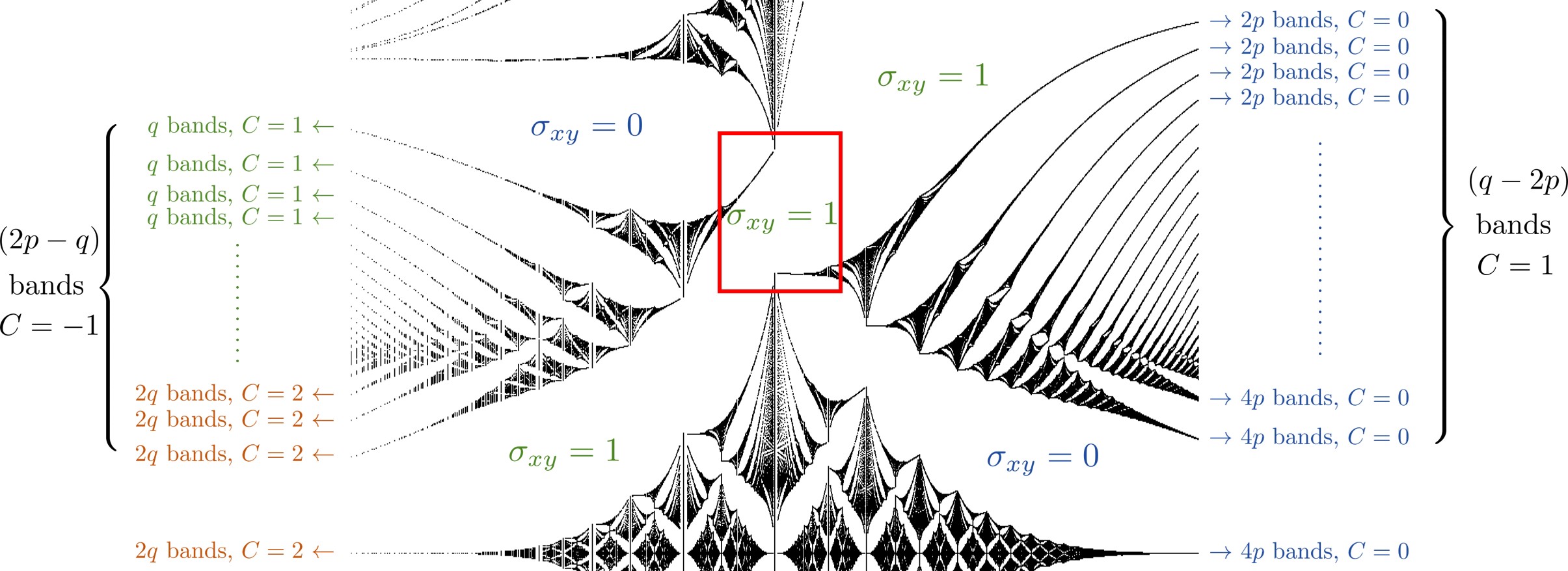}}
\subfigure[]{\label{fig:300}\includegraphics[width=0.25\textwidth]{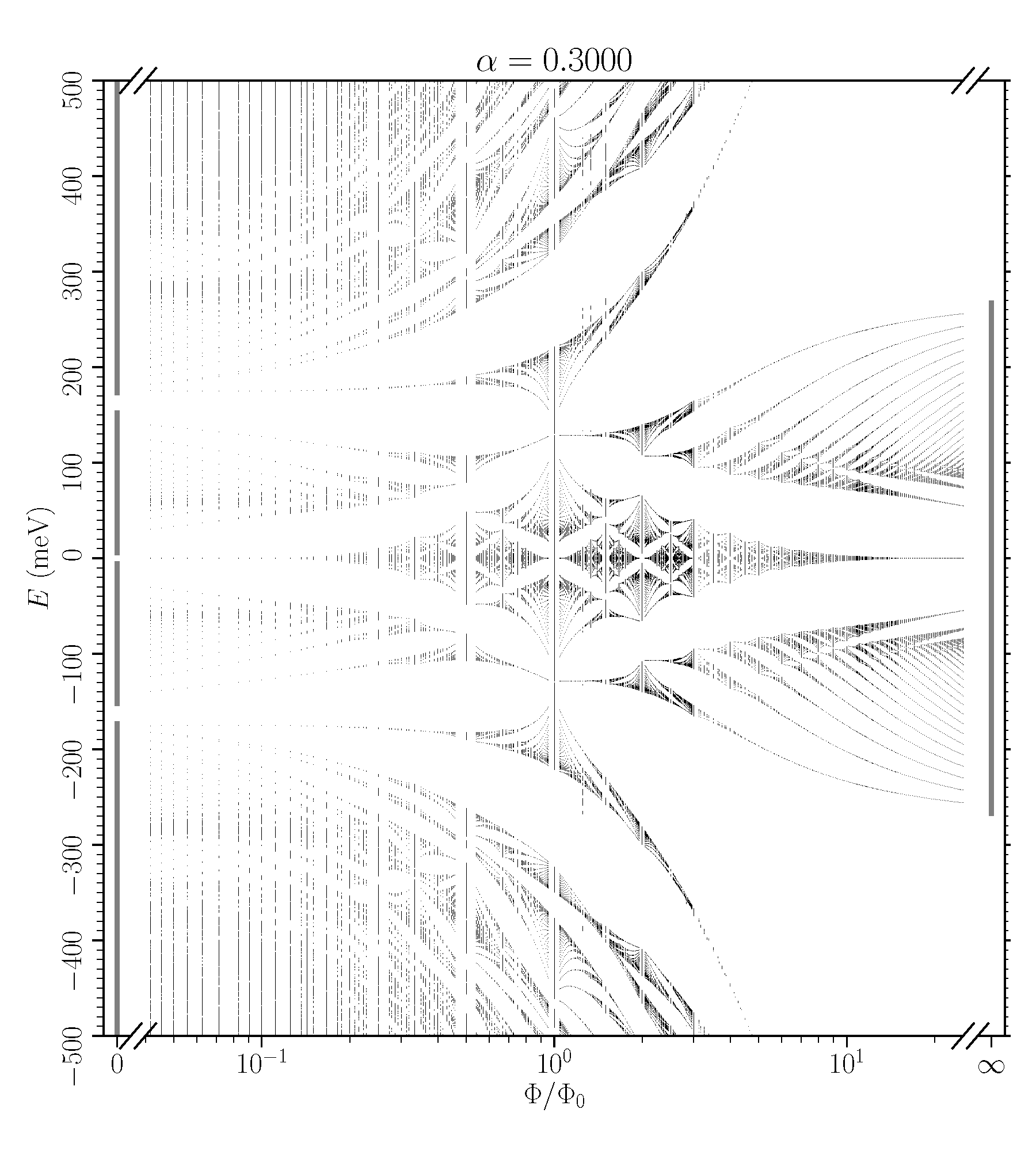}}
\caption{(a) Magnetic bands and their Chern numbers of the butterfly plot at $\alpha = 0.25$. The integer Hall conductivity $\sigma_{xy}$ in unit of $e^2/h$ for the major gaps is given; the $\sigma_{xy}=1$ gap persists adiabatically from small to large fields. Although not shown, the band Chern numbers of the butterfly plot at $\alpha = 0.35$ has the same pattern as labeled here, except for the persisting gap through $\Phi/\Phi_0 =1$ (the gap inside the red box): for $\alpha=0.35$ the $\sigma_{xy}=1$ gap is closed while the $\sigma_{xy}=0$ gap is adiabatically continued (as shown in Fig.~\ref{fig:full350}). (b) Butterfly plot in a wide range of $1/25\leq \Phi/\Phi_0\leq 25$ for $\alpha = 0.30$. Notice that both the major $\sigma_{xy}=0$ and $\sigma_{xy}=1$ gaps are closed at $\Phi/\Phi_0=1$, one can declare that a transition from the former behavior ($\alpha =0.25$) to the latter ($\alpha =0.35$) happens at this value of $\alpha$.}\label{fig:adiabaticity}
\end{figure}

The butterfly plots in a large range of magnetic flux ($\frac{1}{25}\leq \Phi/\Phi_0\leq 25$) for $\alpha=0.25$ and $\alpha=0.35$ are shown in Fig.~\ref{fig:full}. The plots in the small magnetic regime qualitatively agree with the result of Moon et al. \cite{PhysRevB.85.195458} which was obtained at nearby commensurate angles using a tight binding model. By comparing the low energy levels at a flux $\Phi/\Phi_0 = q/2p < 1$ and its reciprocal $\Phi/\Phi_0=2p/q > 1$ in the plots, one notices a resemblance: the butterfly structure at flux $q/2p$ and its reciprocal flux value $2p/q$ look similar. One can establish the details of this correspondence by taking a close look at the low energy magnetic bands, shown in Fig.~\ref{fig:adiabaticity}: the low energy butterfly exhibits a left wing ($\Phi/\Phi_0< 1$) and a right wing ($\Phi/\Phi_0>1$). Near the left (right) edge of the left (right) wing, the magnetic bands collapse to well-defined levels. Analyzing the weight of these levels was our major task, discussed in the main text. At the upper edge, the levels of the left (right) wing each contains $q$ ($2p$) bands, while the levels at the lower edge (note that in the main text these levels are referred to as the LLs in the middle of the active range) of the left (right) wing each contains $2q$ ($4p$) bands.

%Such a mapping is just one manifestation of the duality in the theory of Bloch electrons subject to magnetic field: the quantities at small magnetic field $\Phi/\Phi_0 \ll 1 $ can be mapped to those at large magnetic field $\Phi/\Phi_0\gg 1$ if one defines a fictitious magnetic field $\widetilde{B}$. In the extreme case of the duality between zero magnetic field and infinite magnetic field, the nonmagnetic continuum model \eqref{eq:TBG_no_magnetic} is mapped to a tight binding model (Eq.~\eqref{dualTB}) describing the zeroth Landau level on honeycomb lattice, see App.~\ref{app:infinite_B} for details.

In the main text we have commented on the two different behaviors for the adiabaticity of the band weights: either the weight of the zeroth LLs of the two moir\'{e} DPs (consisting of $2q$ magnetic bands) or the weight of the two nonmagnetic bands (consisting of $4p$ magnetic bands) is adiabataically continued as flux is varied through $\Phi/\Phi_0 = 1$. The former behavior is found at $\alpha = 0.25$ (see Fig.~\ref{fig:adiabaticity}): the gap between the active $2q$ magnetic bands (zeroth LLs of the two moir\'{e} DPs) and higher bands persists at all flux values, while the gap between $4p$ magnetic bands in the active range and the the remote bands closes at $\Phi/\Phi_0=1$; the latter behavior is found at $\alpha=0.35$, where the persisting gap and the closing gap are switched. Note that there is a gap at CNP (although the two bands above and below it can also touch at this point), which corresponds to having a vanishing Hall conductivity, $\sigma_{xy}=0$; the Hall conductivity (in the units of $e^2/h$) for other gaps can then be obtained by suming over all the Chern numbers for the bands below this gap but above CNP.

In this way, we find that the gap above CNP persisting from small to large fields has a unit Hall conductivity, $\sigma_{xy}=1$ for $\alpha=0.25$, and $\sigma_{xy} = 0$ for $\alpha = 0.35$ (see Fig.~\ref{fig:250analysis}). Furthermore, we found that a transition between these two behaviors happens at $\alpha\approx 0.30$, where both gaps are closed at $\Phi/\Phi_0=1$, see Fig.~\ref{fig:300}. The difference in the adiabaticity behavior found above may have observable effects in quantum Hall experiments.
%: the quantized Hall conductivity $\sigma_{xy}$ is proportional to the Chern number of the gap at which fermi surface lies. 

Fig.~\ref{fig:250analysis} shows the Chern numbers of the LLs lying between the two major gaps (the gap with $\sigma_{xy}=0$ and the gap with $\sigma_{xy}=1$). Notice that at small magnetic field limit, the total Chern number of magnetic bands within the active range vanishes, while for large magnetic field limit the total Chern number for the $2q$ bands is 2. The results are in accordance with a computation based on Streda's formula.

\section{Wannier plot}\label{app:wannier}

In this section we give a prescription for extracting the weight information in the butterfly plot. The final result is the wannier plots shown in Fig.~\ref{fig:wannier} in the main text.

The wannier plot is a density plot, which shows the density of states $\rho(n,B)$ as a function of carrier density $n$ and magnetic field $B$. In principle it can be obtained by transcribing the energy spectrum (butterfly) plot according to the following method:

\begin{itemize}
\item The density of states are obtained by broadening the $\delta$ functions using a Cauchy distribution:
\begin{equation}\label{eq1}
\rho(E) = \sum\limits_i \delta(E-E_i)\rightarrow \sum_i \frac{1}{\pi} 
\frac{\gamma}{(E-E_i)^2+\gamma^2},
\end{equation}
where the parameter $\gamma$ is an empirical parameter. This parameter is adjusted for each value of $\alpha$ to achieve optimal resolution for the Landau levels.
\item The carrier density at energy $E$ is obtained by integrating density of states from above:
\begin{equation}\label{eq2}
n(E) = \sum\limits_i \theta(E-E_i) \rightarrow \sum_i \frac{1}{\pi}\arctan\left(\frac{E-E_i}{\gamma}\right),
\end{equation}
\end{itemize}

Due to the fact that the number of magnetic bands scales with $p$, where $p$ and $q$ are coprime numbers satisfying $\Phi/\Phi_0 = q/2p$ (and also noting the prescription for finding densities given in the main text), a normalization factor $\frac{1}{2p}$ has to be given to the expressions Eq.~\eqref{eq1} and \eqref{eq2} when the density of a numerically found magnetic band is calculated. This guarantees that the active range corresponds to $n\in[-1,1]$ in a Wannier plot. 

As mentioned in Fig.~\ref{fig:wannier}, the colors therein in fact correspond to a rescaled density of states $\rho/\rho_{\text{max}}$, where $\rho_{\text{max}}$ is a large value of density of states which sets the rightmost scale of the colorbar. The value of $\rho_{\text{max}}$ is different for the three subplots; furthermore it may not be the actual largest density of states computed from the butterfly plot. $\rho_{\text{max}}$ is chosen simply to obtain the best resolution of the LLs for the figures.

\section{Semiclassical energies}\label{app:semiclassics}
\begin{figure*}[!thb]
\centering
 \subfigure[]{\label{fig:butterfly_comparison_a} \includegraphics[width=.32\textwidth]{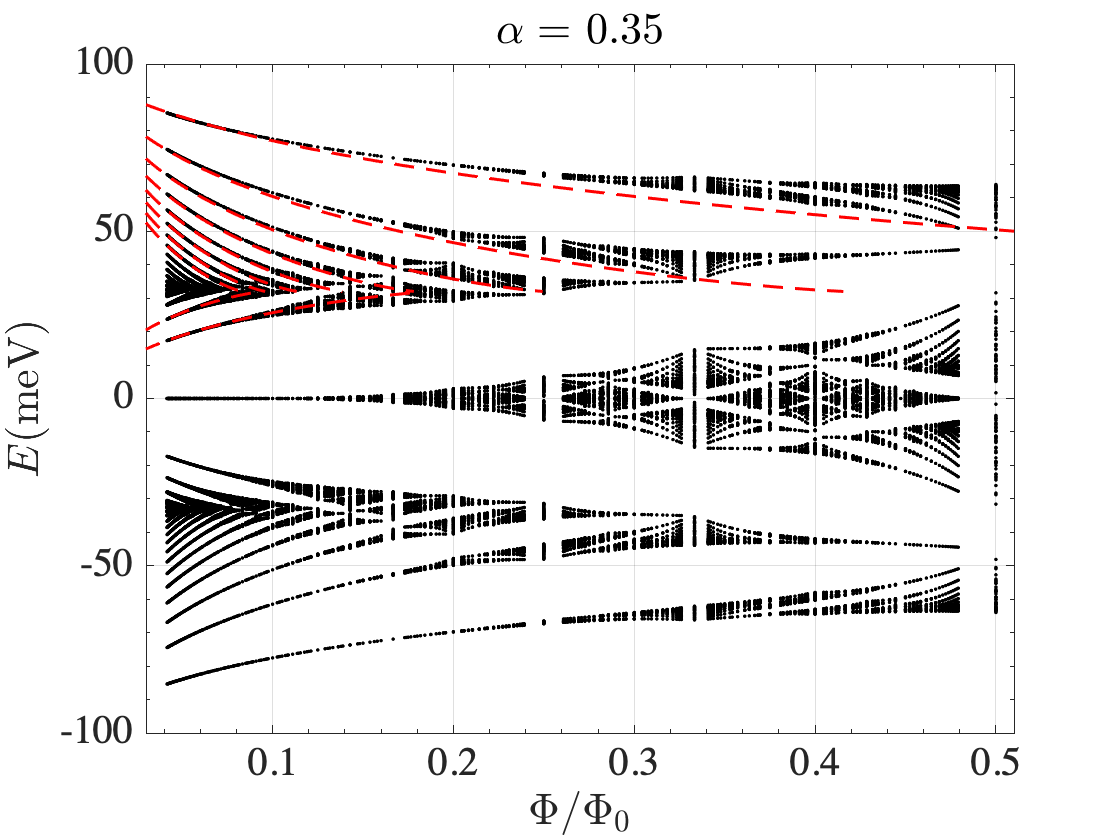}} 
 \subfigure[]{\label{fig:butterfly_comparison_b} \includegraphics[width=.32\textwidth]{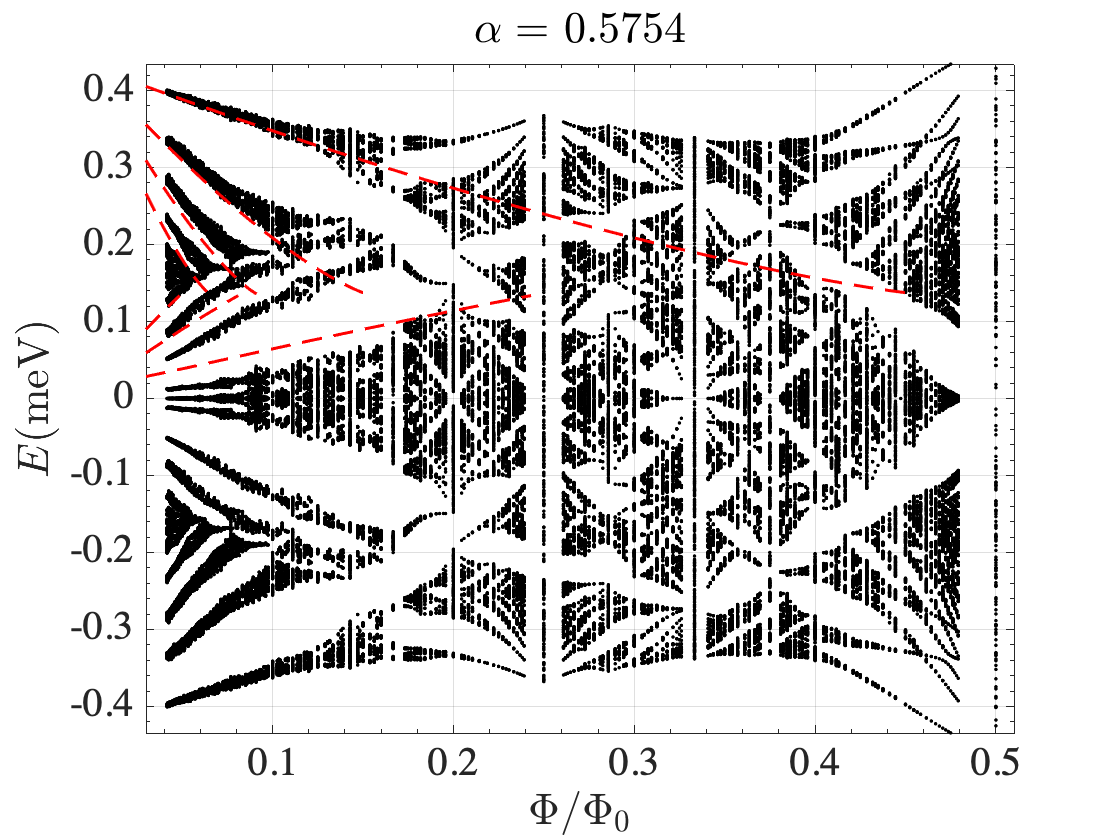}}
 \subfigure[]{\label{fig:butterfly_comparison_c}  \includegraphics[width=.32\textwidth]{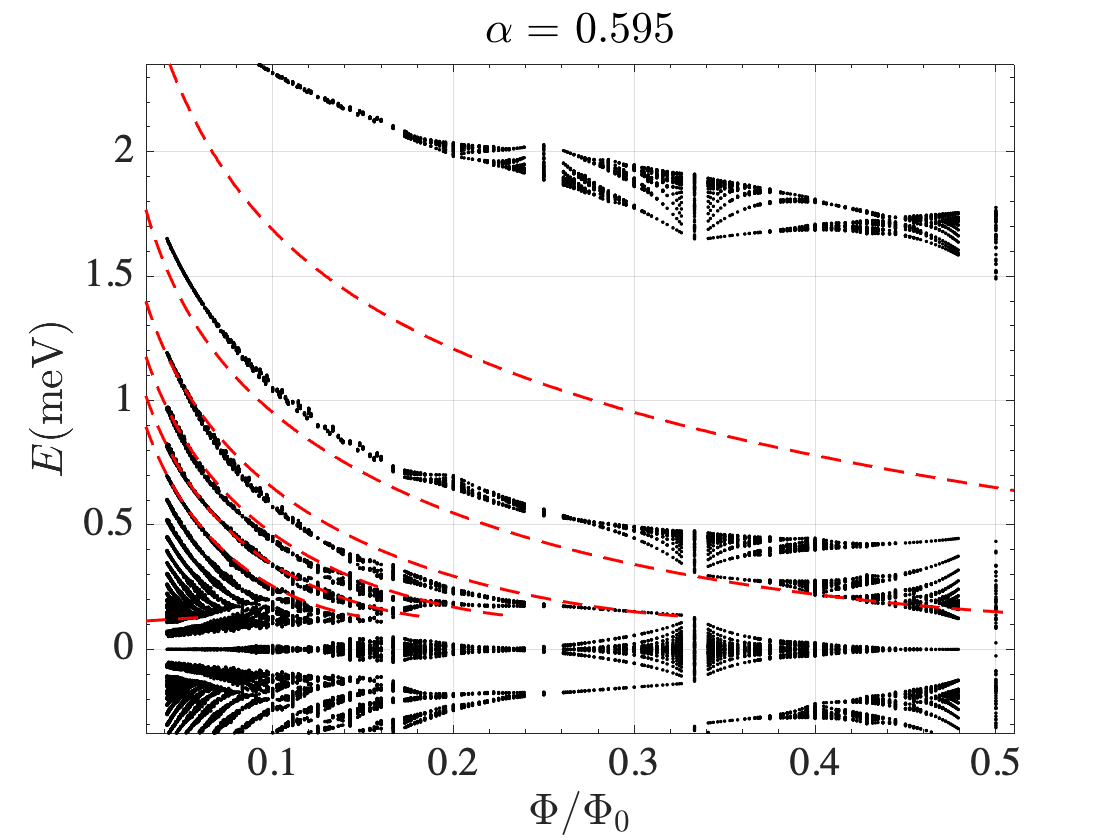}}
 \caption{ \label{fig:butterfly_comparison} A comparison between LLs found using semiclassical analysis and those in butterfly plots. The semiclassical energies are shown by red dashed lines on top of the butterfly plots.} 
\end{figure*}

In this section of the supplemental information, we compare the energies found using the semiclassical analysis and the energy levels in the butterfly plots at small (relevant to experiments) fields. The semiclassical energies are shown as red dashed lines on top of butterfly plots in Fig.~\ref{fig:butterfly_comparison}. Each semiclassical line is found by imposing the relevant quantization condition on the enclosed area in different areas of the BZ; the lines are continued until the energy reaches the saddle contour energy (see main text). 

The agreement between the two sets of energies is best at $\alpha = 0.35$, however, within the magic range the agreement is less pronounced; at $\alpha = 0.5754$, the agreement of the edge levels is better while in the middle of the active range the LLs in the butterfly plot have a smaller energy than those found by semiclassical analysis. This discrepancy can be understood by noting that at this value of $\alpha$ the two active bands approach each other at the $\Gamma$ point which can lead to interband mixing, and that we are neglecting this here. Still good qualitative agreement can be seen. At $\alpha = 0.595$ finally, the edge LLs found by the two methods show better agreement as one gets further from the edge; and in the middle of the active range, the only 3-fold degenerate LL found in the butterfly plot is acceptably close to the 3-fold LL found by semiclassical analysis.

\section{Broken particle-hole symmetry}\label{app:ph_broken}
\begin{figure*}[!thb]
\centering
 \subfigure[]{\label{fig:nonmagnetic_broken_ph_a} \includegraphics[width=.4\textwidth]{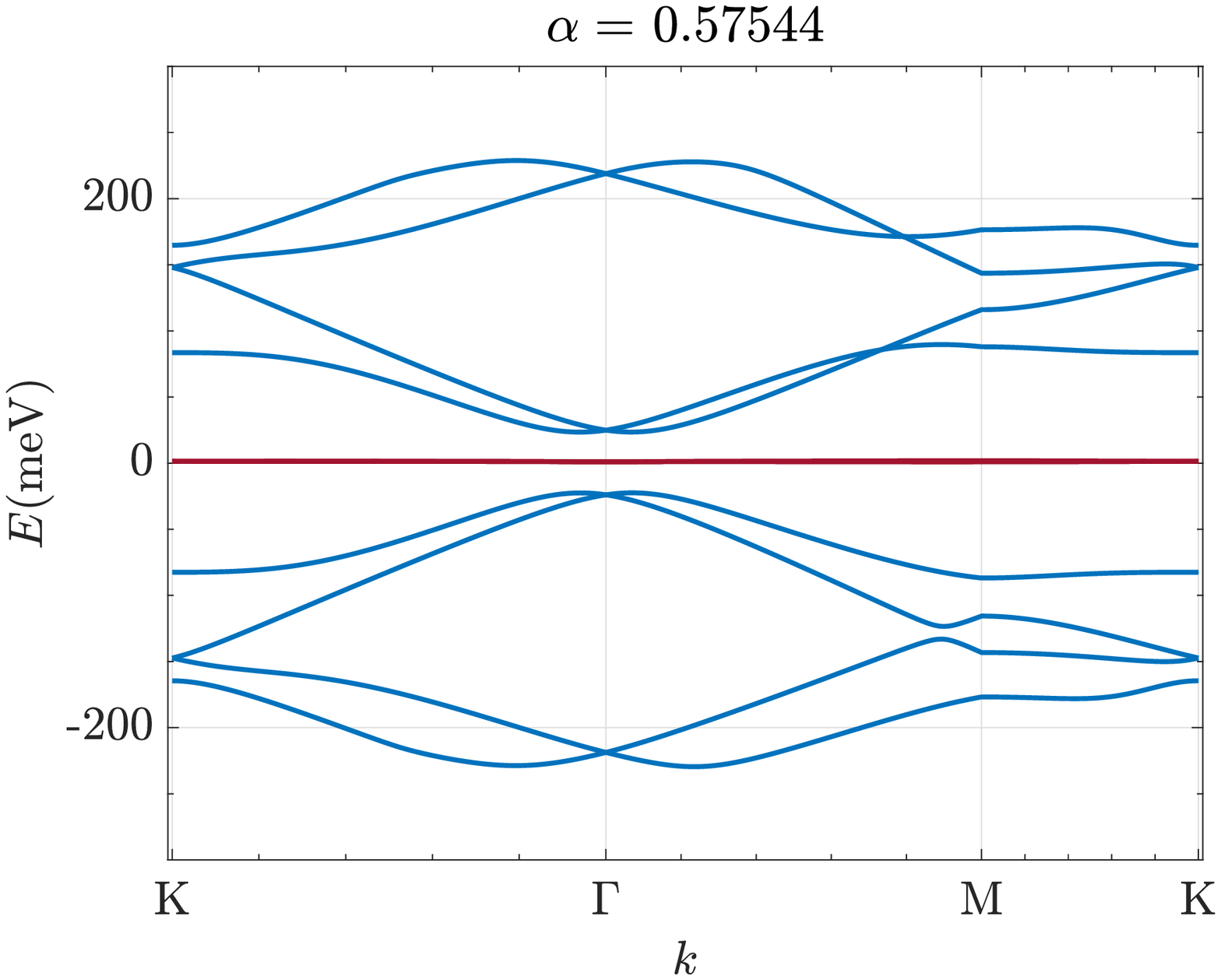}} 
 \subfigure[]{\label{fig:nonmagnetic_broken_ph_b} \includegraphics[width=.4\textwidth]{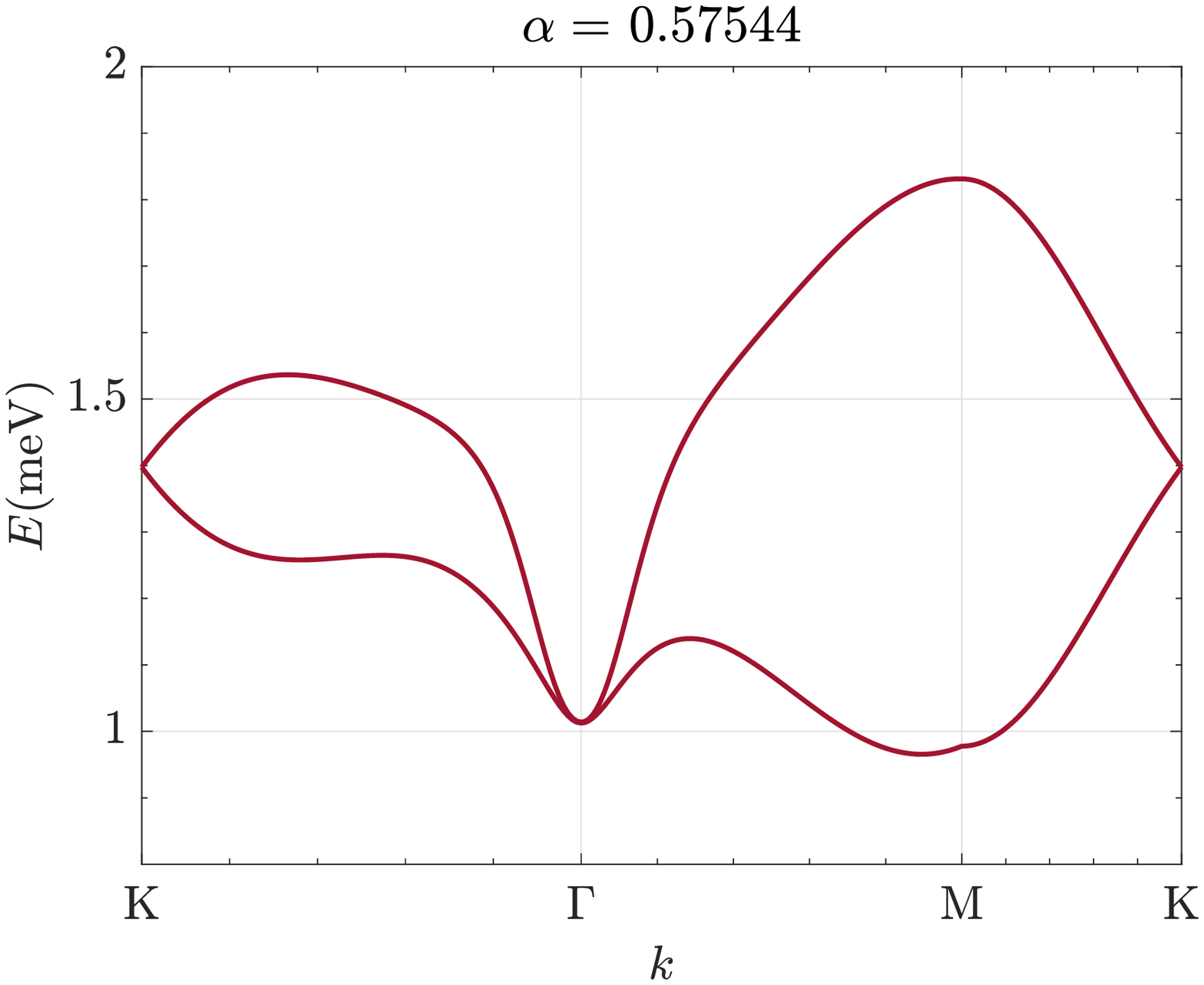}}
 \caption{\label{fig:nonmagnetic_broken_ph} Nonmagnetic bands close to CNP, for $\eta = 0.82$ and $\alpha = \alpha_1$. The ten closest bands and the two active bands are shown in (a) and (b) respectively. At this value of $\alpha$ the active bands show a considerable asymmetry.} 
\end{figure*}

\begin{figure}[!thb]
\centering
\includegraphics[width=0.7\textwidth]{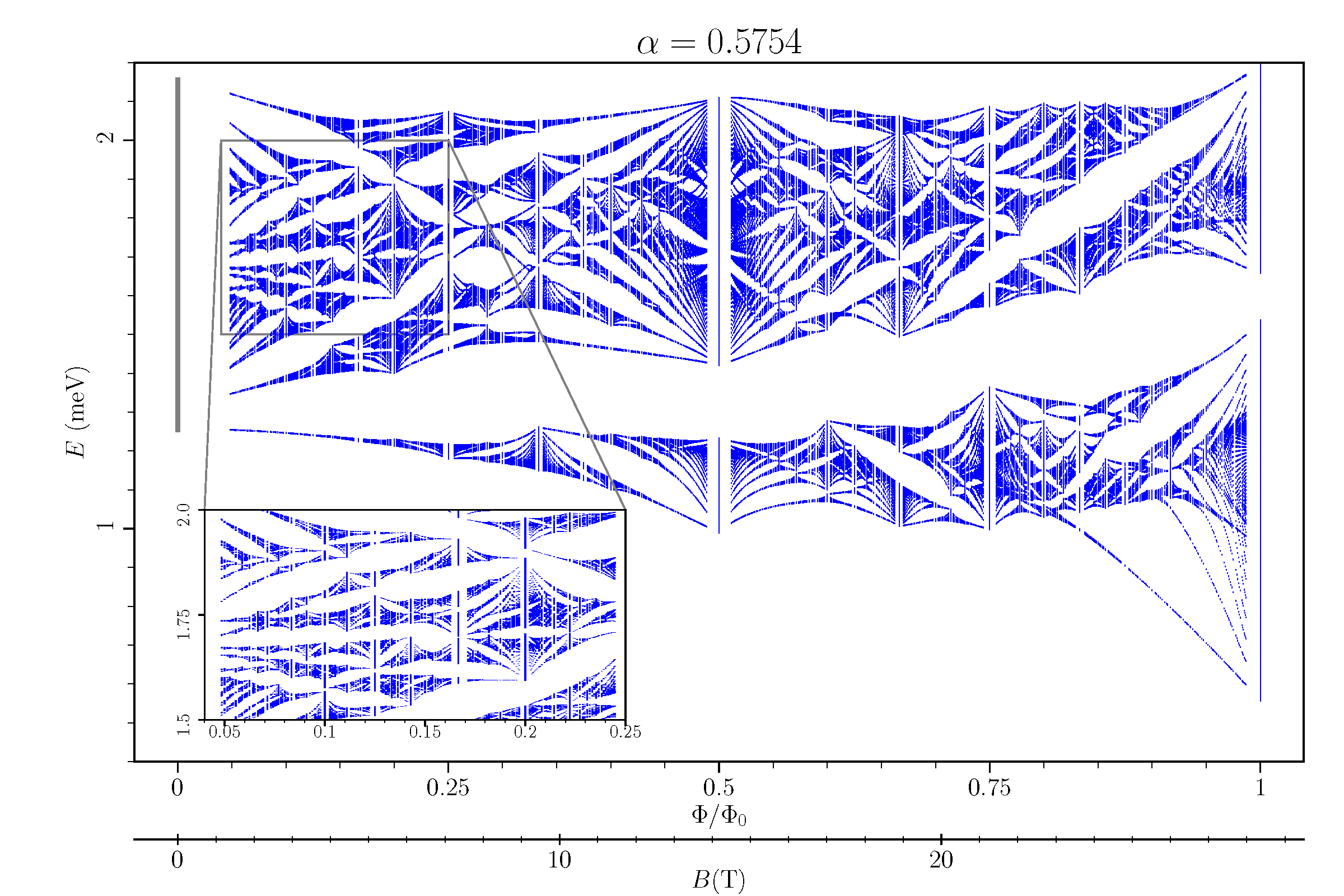}
\caption{\label{fig:butterfly_broken_ph}Butterfly plot for the $(\eta,\mathbf{R})=(0.82,\mathbf{R}_{\tau^z\theta/2})$ model (i.e. both lattice corrugation and sublattice pseudospin considered) at $\alpha = 0.5754$. The inset shows magnified version of the magnetic energy levels at small fields.}\label{fig:5754_thetadev_1}
\end{figure}

As is discussed in the main text, with the inclusion of sublattice rotation matrices $\mathbf{R}_{\tau^z\theta/2}$ the particle-hole symmetry in the magnetic spectrum is broken. This does not have a major effect on the filling factor sequences discussed in the main text when the twist angle is outside the magic range. However, close to $\alpha = \alpha_1$ (where the two active bands have a quadratic band touching), the situation can be different; at $\alpha_1$ the nonmagnetic active bands touch each other below the energy of DPs as showin in Fig.~\ref{fig:nonmagnetic_broken_ph}. As a result of this, the orbits formed in the top band around the $\Gamma$ point can have lower energy than the zero energy LLs of moir\'e DPs, when $\alpha$ is close to $\alpha_1$. As can be seen in Fig.~\ref{fig:butterfly_broken_ph}, for small magnetic field level crossings in the middle of the active range can occur which can potentially result in an abrupt change of the filling factor sequence.

\section{Mapping the model in the infinite magnetic field limit to a zero field model}
\label{app:infinite_B}

Although the magnetic model introduced in App.~\ref{app:the_model} looks rather complicated, it can be simplified in the inifinte magnetic field limit $B\rightarrow \infty$. This is due to the decoupling of the bare LLs, which can be observed in the butterfly diagram: at infinite magnetic field limit $\Phi/\Phi_0=q/2p\rightarrow \infty$, the low energy part of the butterfly spectrum consists of $2q$ magnetic bands, which are precisely the two zeroth LLs of the Dirac points of the two graphene sheets, i.e. the zero energy solution of the part $H_{\text{LL}}$ which are widened due to moir\'e lattice. They are infinitely apart in energy from the other LLs. Furthermore, as magnetic field approaches infinity, the band width of each of the $2q$ magnetic bands tends to vanish, and eventually these $2q$ bands merge into one continuous band. 

As discussed in App.~\ref{app:duality_adiabaticity}, the butterfly plot (see Fig.~\ref{fig:full}) shows a kind of duality between small and large fluxes: $2q$ magnetic bands appear at charge neutrality point for small fields and eventually evolve (with the possibility for a gap closing) into the zeroth LLs at large fields; on the other hand, the $4p$ bands corresponding to the total weight of the two active bands at small fields also evolve (again with the possibility for a gap closing) and gradually converge to charge neutrality point at large fields. This spectral duality is reminiscent of the duality of the Harper's equation in the weak potential limit and the strong potential limit \cite{thouless1982quantized}, where the flux quantization condition of one case is the inverse of the other.

These facts motivate us to look for a simple description of the $2q$ bands at large magnetic fields. Such a $2q$ by $2q$ Hamiltonian $\mathcal{H}_{2q\times 2q}$ can be easily obtained by projecting the Hamiltonian introduced in App.~\ref{app:the_model} onto the $2q$-fold basis of the zeroth LLs. Crucially, $\mathcal{H}_{2q\times 2q}$ can be interpreted differently as a fictitious Hofstadter Hamiltonian obtained from a \emph{tight binding model} on a honeycomb lattice subject to a commensurate \emph{dual} magnetic field $\widetilde{B}$. Here the tight binding model and the dual magnetic field $\widetilde{B}$ are fictitious; the sublattice index $s=\widetilde{1},\widetilde{2}$ of the fictitious honeycomb lattice in fact is dual to the zeroth LLs of layers 1 and 2, and the fictitious magnetic field $\widetilde{B}$ requires that the dual flux per honeycomb unit cell $\widetilde{\Phi}$ in this dual magnetic problem is the inverse of the physical flux per unit cell $\widetilde{\Phi}/\Phi_0 = \Phi_0/\Phi = 2p/q$. From now on we will refer to this fictitious theory as the dual theory for simplicity. 

To be more precise, we introduce the fictitious honeycomb lattice in the dual theory by specifying the vectors $\bm{a}_{0,1,2}$ from a $\widetilde{1}$ sublattice site to its three nearest neighbor (NN) $\widetilde{2}$ sites: 
\begin{equation}
\bm{a}_0 = (\widetilde{a},0),\qquad \bm{a}_1 = (-\widetilde{a}/2,\widetilde{b}),\quad\text{and}\quad\bm{a}_2 = (-\widetilde{a}/2,-\widetilde{b}),
\end{equation}
where we defined 
\begin{equation}\label{eq:defab}
\widetilde{a} = k_\theta \ell_{\text{B}}^2\quad\text{and}\quad \widetilde{b} = \Delta.
\end{equation}
Depending on the ratio between $\widetilde{a}$ and $\widetilde{b}$ the honeycomb may appear elongated or compressed but this will not affect the physics we study. The tight binding Hamiltonian we propose in the dual theory is the NN hopping Hamiltonian on the honeycomb lattice: 
\begin{equation}\label{eq:honeycombdual}
H = -t \sum_{i}\sum_{j=0,1,2} c^\dag_{\bm{r}_i,\widetilde{1}} c_{\bm{r}_i+\bm{a}_j,\widetilde{2}} + \text{H.c.},
\end{equation}
where $i$ runs over all $\widetilde{1}$ sublattice sites, and the hopping amplitude $t = \alpha \eta \hbar v_{\text{F}} k_\theta$. Let us define lattice translation vectors $\bm{t}_1 =  (3\widetilde{a}/2,-\widetilde{b})$ and $\bm{t}_2 = (3\widetilde{a}/2,\widetilde{b})$ for the $\widetilde{1}$ sublattice. 
%The position vector of every $\widetilde{1}$ site can be written as $\bm{r}_i=m_i\bm{t}_1+n_i\bm{t}_2$, where $(m_i,n_i)$ are either both odd or even numbers.

Then, we apply the fictitious magnetic field $\widetilde{B}$ by using the Landau gauge $\widetilde{\bm{A}} = (-y\widetilde{B},0)$ ($y$ is continuous here). The magnetic Hamiltonian $\widetilde{H}'$ can be obtained via a Peierl's substitution:
\begin{equation}\label{dualTB}
\widetilde{H}'
 = -t \sum_i e^{\frac{2\pi i}{\Phi_0}\widetilde{B}\widetilde{a}\widetilde{b}y_i} c^\dag_{\bm{r}_i,\widetilde{1}}c_{\bm{r}_i+\bm{a}_0,\widetilde{2}}
+ e^{-\frac{2\pi i}{\Phi_0}\frac{\widetilde{B}\widetilde{a}\widetilde{b}}{2}(y_i+1/2)} c^\dag_{\bm{r}_i,\widetilde{1}}c_{\bm{r}_i+\bm{a}_1,\widetilde{2}}
+ e^{-\frac{2\pi i}{\Phi_0}\frac{\widetilde{B}\widetilde{a}\widetilde{b}}{2}(y_i-1/2)} c^\dag_{\bm{r}_i,\widetilde{1}}c_{\bm{r}_i+\bm{a}_2,\widetilde{2}} + \text{H.c.},
\end{equation}
note that $y_i$ is an integer. Next, we perform a gauge transformation
\begin{equation}
c_{\bm{r}_i,\widetilde{1}}\rightarrow e^{\frac{\pi i}{\Phi_0}\widetilde{B}\widetilde{a}\widetilde{b} y_i}c_{\bm{r}_i,\widetilde{1}},\qquad
c_{\bm{r}_i+\bm{a}_0,\widetilde{2}}\rightarrow
e^{-\frac{\pi i}{\Phi_0}\widetilde{B}\widetilde{a}\widetilde{b} y_i}c_{\bm{r}_i+\bm{a}_0,\widetilde{2}},
\end{equation}
under which the Hamiltonian transforms as $\widetilde{H}'\rightarrow \widetilde{H}$, where
\begin{equation}
\widetilde{H}=
-t \sum_i c^\dag_{\bm{r}_i,\widetilde{1}}c_{\bm{r}_i+\bm{a}_0,\widetilde{2}}
+ e^{-\frac{2\pi i}{\Phi_0}\frac{\widetilde{\Phi}}{2}(y_i+1/2)} c^\dag_{\bm{r}_i,\widetilde{1}}c_{\bm{r}_i+\bm{a}_1,\widetilde{2}}
+ e^{-\frac{2\pi i}{\Phi_0}\frac{\widetilde{\Phi}}{2}(y_i-1/2)} c^\dag_{\bm{r}_i,\widetilde{1}}c_{\bm{r}_i+\bm{a}_2,\widetilde{2}} + \text{H.c.}.
\end{equation}
Note that we have defined flux per unit cell $\widetilde{\Phi} = \widetilde{B}\widetilde{\mathcal{A}}$, where $\widetilde{\mathcal{A}}=3\widetilde{a}\widetilde{b}$ is the hexagonal unit cell area of the dual honeycomb lattice. Having the purpose of getting a dual theory to the initial model, we impose the following commensurate flux condition
\begin{equation}
\frac{\widetilde{\Phi}}{\Phi_0} = \frac{2p}{q},
\end{equation}
under this condition the Hamiltonian $\widetilde{H}$ has new translation symmetry along vectors $\widetilde{\bm{t}}_1=(3\widetilde{a}/2,-q\widetilde{b})$, $\widetilde{\bm{t}}_2 = (3\widetilde{a}/2,q\widetilde{b})$ for odd $q$, or $\widetilde{\bm{t}}_1 = (3\widetilde{a},0)$, $\widetilde{\bm{t}}_2 = (0,q\widetilde{b})$ for even $q$. 
Next we turn to using the Fourier transformed operators $c_{\widetilde{\bm{k}},l_i,s}$ instead of $c_{\bm{r}_i,s}$, where $l_i \equiv y_i \ \mathrm{mod} \ q$.
%one can write $c_{\bm{r}_i,\widetilde{1}}= e^{i \widetilde{\bm{k}}\cdot \bm{r}_i} c_{j_i,\widetilde{1}}$ and $c_{\bm{r}_i+\bm{a}_0,\widetilde{2}} = e^{i \widetilde{\bm{k}}\cdot \bm{r}_i}c_{j_i,\widetilde{2}}$, where $j_i\equiv n_i$ mod $q$ is the remainder of $n_i$ divided by $q$;
 $\widetilde{\bm{k}}$ belongs to the dual magnetic BZ (which is valid for $q$ being both odd and even)
\begin{equation}\label{eq:dualBZ}
\widetilde{\bm{k}} = (\widetilde{k}_1,\widetilde{k}_2) \in \left[0,\frac{2\pi}{3\widetilde{a}}\right]\times\left[0,\frac{2\pi}{q\widetilde{b}}\right] = \left[0,\frac{2\pi}{3k_\theta \ell_{\text{B}}^2}\right]\times\left[0,\frac{2\pi}{q \Delta}\right],
\end{equation}
And the Hamiltonian becomes
\begin{equation}\label{eq:honeycomb_hofstadter}
\widetilde{H}(\widetilde{\bm{k}})
=
-t\sum_{l=0}^{q-1}
c^\dag_{l,\widetilde{1}}c_{l,\widetilde{2}} + e^{-i\widetilde{\bm{k}}\cdot\bm{t}_1}e^{-2\pi i \frac{\widetilde{\Phi}}{\Phi_0}(l+\frac{1}{2})} c^\dag_{l,\widetilde{1}}c_{l+1,\widetilde{2}}
+e^{-i\widetilde{\bm{k}}\cdot\bm{t}_2}e^{-2\pi i \frac{\widetilde{\Phi}}{\Phi_0}(l-\frac{1}{2})} c^\dag_{l,\widetilde{1}}c_{l-1,\widetilde{2}} + \text{H.c.},
\end{equation}
this Hamiltonian, written in basis of $c_{l,\widetilde{1}/\widetilde{2}}$, will give a $2q$ by $2q$ hermitian matrix $\widetilde{\mathcal{H}}_{2q\times2q}$. Comparing this with the matrix elements given in Eq.~\eqref{eq:tunnelingterms}, we establish the following relations between the corresponding quantities in the original theory and the dual theory:
\begin{itemize}
\item $\widetilde{\mathcal{H}}_{2q\times2q}$ is equal to $\mathcal{H}_{2q\times2q}$, if one relates $\widetilde{k}$ to the magnetic momentum of the original magnetic theory by \begin{equation}\label{eq:rescalingk}
\widetilde{k}_x = k_x\quad\text{and}\quad \widetilde{k}_y = -k_y;
\end{equation}
\item The dual magnetic flux is related to the original magnetic flux by \begin{equation}\label{eq:phiphi}
\frac{\widetilde{\Phi}}{\Phi_0}=\frac{\Phi_0}{\Phi} = \frac{2p}{q};
\end{equation}
\item From Eq.~\eqref{eq:phiphi}, the magnetic field of the original theory and the dual theory are actually equal:
\begin{equation}
B = \widetilde{B}.
\end{equation}
This implies that in the large magnetic field limit $B\propto q/p\rightarrow \infty$, the dual magnetic field $\widetilde{B}$ is also large. However the dual magnetic flux per unit cell, $\widetilde{\Phi}$, approaches zero in this limit, despite $\widetilde{B}$ being large. This is due to the fact that, by definition, the dual unit cell area $\widetilde{A}\propto (p/q)^2$, while $\widetilde{B}\propto q/p$.
\end{itemize}

\begin{figure}[!thb]
\centering
\includegraphics[width=0.8\textwidth]{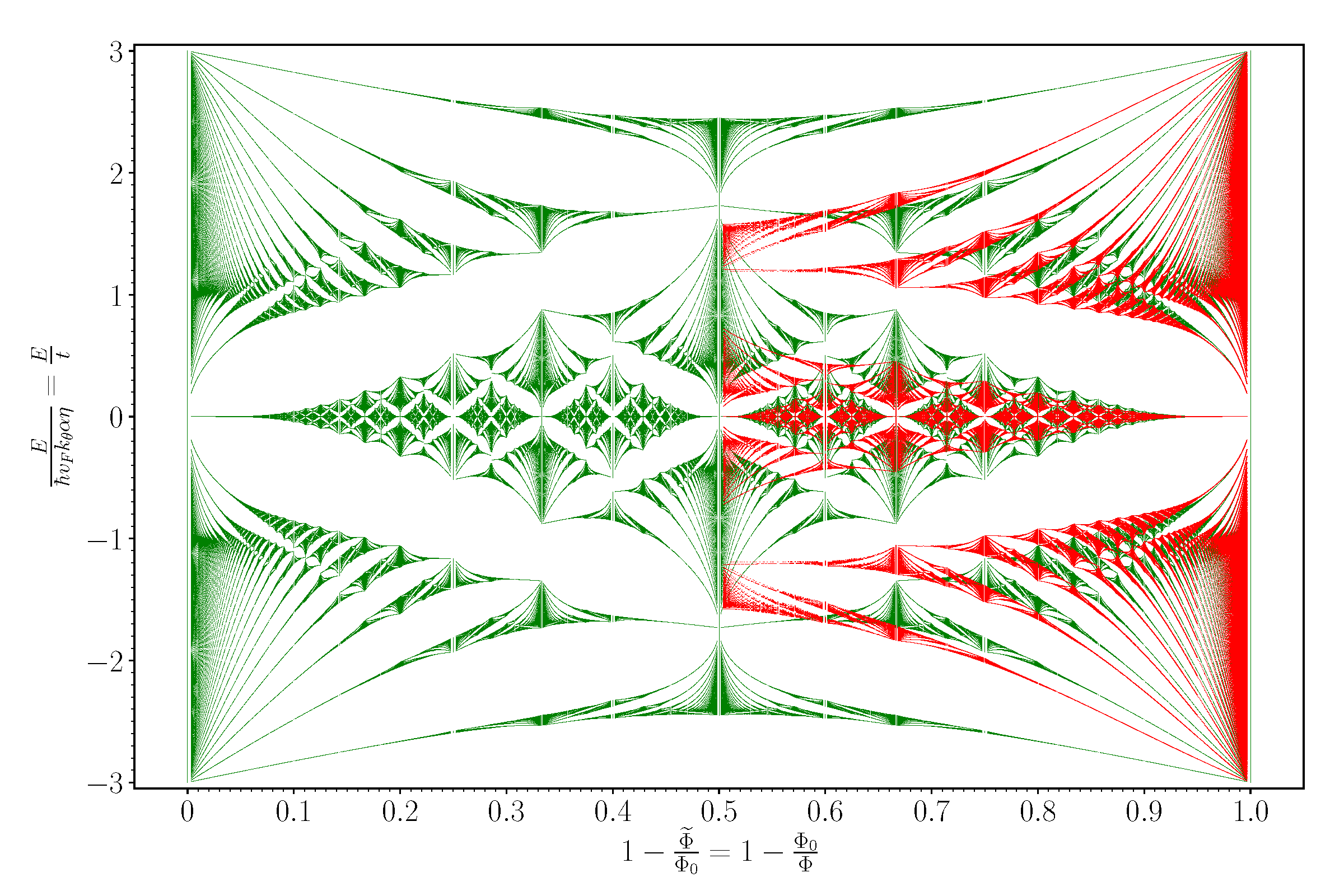}
\caption{Comparison of the butterfly plots obtained from the original theory and the dual theory. Red: the butterfly spectrum of the model in App.~\ref{app:the_model} for $\alpha=0.3$ in the flux range $\frac{1}{2}\leq 1-\Phi_0/\Phi\leq 1$; Green: the butterfly spectrum of the dual model (see Eq.~\eqref{eq:honeycomb_hofstadter}) -- honeycomb Hofstadter butterfly. 
	Note that the green butterfly does not change if one uses $\widetilde{\Phi}/\Phi_0$ as the horizontal axis due to its mirror symmetry about  $\widetilde{\Phi}/\Phi_0 = 1/2$. 
	In the limit $\widetilde{\Phi}/\Phi_0 = \Phi_0/\Phi\rightarrow 0$ (i.e. the rightmost part of the plot), the spectra of the two models become identical.}\label{fig:honeycomb_butterfly}
\end{figure}

% In this dual language, such a tight binding model (in absence of the field $\widetilde{B}$) can be Fourier transformed to arrive at a simple Bloch Hamiltonian which resembles the standard nearest neighbor monolayer graphene Hamiltonian:
%\begin{gather}\label{dualH}
%\widetilde{H}(\bm{k})
%=
%\alpha \eta \left(\begin{array}{cc} 0 & f(\bm{k}) \\ f^*(\bm{k}) & 0 \end{array}\right),\nonumber\\
%f(\bm{k}) = 1 + e^{-3 i k_\theta y_0/2}\, e^{-i \, k_2 \Delta } + 
%e^{-3 i k_\theta y_0/2}\, e^{i \, k_2 \Delta },
%\end{gather}
%and we choose a rectangular Brilloun zone for later use
%\begin{equation}\label{eq:dualBZ}
%0 < \frac{3}{2} k_1 k_\theta \ell_{\text{B}}^2 = \frac{3}{2}k_\theta y_0 < 2\pi, \qquad \quad 0 < 2k_2 \Delta< 2\pi.
%\end{equation}
Therefore, we have established a duality map between the model of App.~\ref{app:the_model} in the large magnetic flux limit and a dual model describing ``monolayer graphene'' in small (fictitious) magnetic flux limit. A couple of observations immediately follow from this duality:
\begin{itemize}
\item The bandwidth of the two zeroth LLs at infinite magnetic field is equal to that of the tight binding model: $-3  \leq  \frac{1}{t} E \leq 3$, in agreement with numerical results (see Fig.~\ref{fig:adiabaticity}).
\item At large flux $\Phi/\Phi_0\gg 1$, the butterfly plot of the $2q$ magnetic bands of the two zeroth LLs as a function of the inversed flux $\Phi_0/\Phi$ should be similar to the Hofstadter butterfly plot of the honeycomb lattice \cite{rammal1985landau}, see Fig.~\ref{fig:honeycomb_butterfly}.
\item Since the butterfly at extreme fields admits two physical pictures (in the original theory and dual theory), it is interesting to compute the band Chern numbers in both theories, and try to understand their relations. Regarding this the first claim is that the for the same band in the spectrum, band Chern number may not be equal in the two theories. As shown above, the Chern numbers of the group of bands near the lower or upper edges of the original model in the large magnetic field always vanish, while the fictitious Chern numbers of the bands calculated in the fictitious tight binding model near the upper and lower edges have the values $1$ and $2$ respectively \cite{PhysRevB.74.205414,Agazzi2014}. The correspondence between the Chern numbers in the two theories will be established in App.~\ref{app:chern}.
\end{itemize}

\section{Topological transitions of the non-magnetic continuum model}\label{app:nonmagnetic}

In this section we give a complete description of the topological transitions happening in the model Hamiltonian for zero magnetic field \eqref{eq:TBG_no_magnetic}, as the twisting angle $\theta$ (or the parameter $\alpha$) is varied in the magic angle range. Recall that in this model there are the lattice corrugation parameter $\eta$ and rotation matrix $\mathbf{R}$ for sublattice pseudospin $\sigma$, and in the main text we have taken a physical value $\eta = 0.82$\cite{koshino2018maximally} and neglected the effect of sublattice pseudospin rotation $\mathbf{R}$. Here we will use the symbol $\mathbf{R} = 1_{2\times 2}$ for the case of neglecting this rotation, and $\mathbf{R} = \mathbf{R}_{\tau^z\theta/2}$ for the case otherwise. Note the latter case introduces a small particle-hole asymmetry into the model.

\begin{figure}[thb!]
    \centering
 \subfigure[]{\label{fig:nonBa}\includegraphics[width=0.38\textwidth]{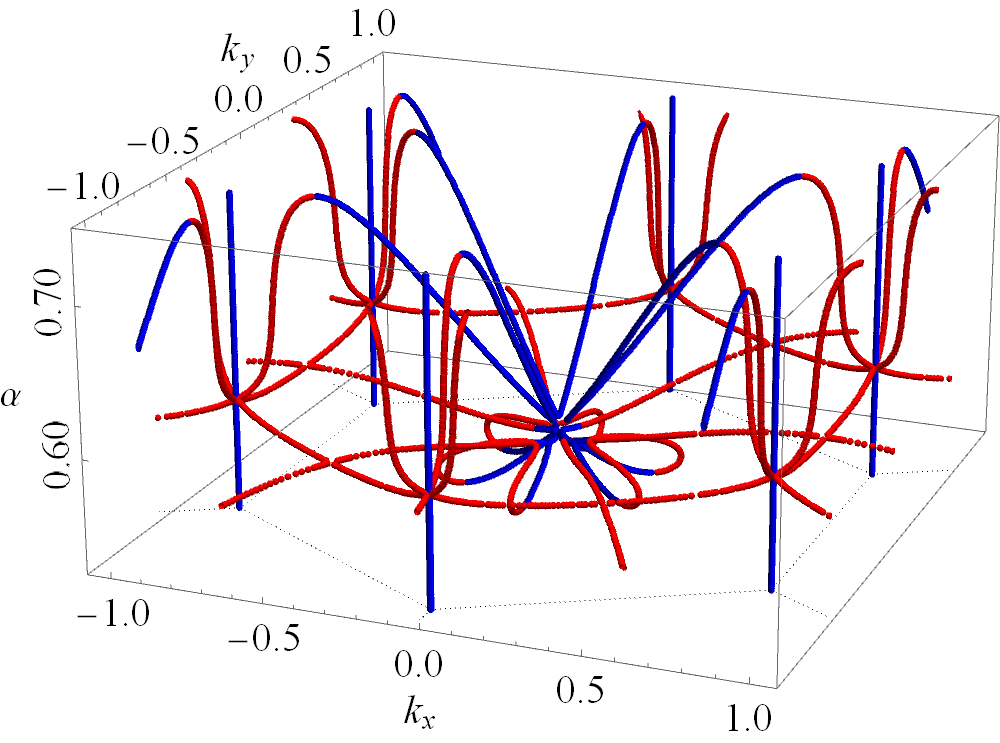}}
 \hspace{0.13in} 
\subfigure[]{\label{fig:nonBb}\includegraphics[width=0.44\textwidth]{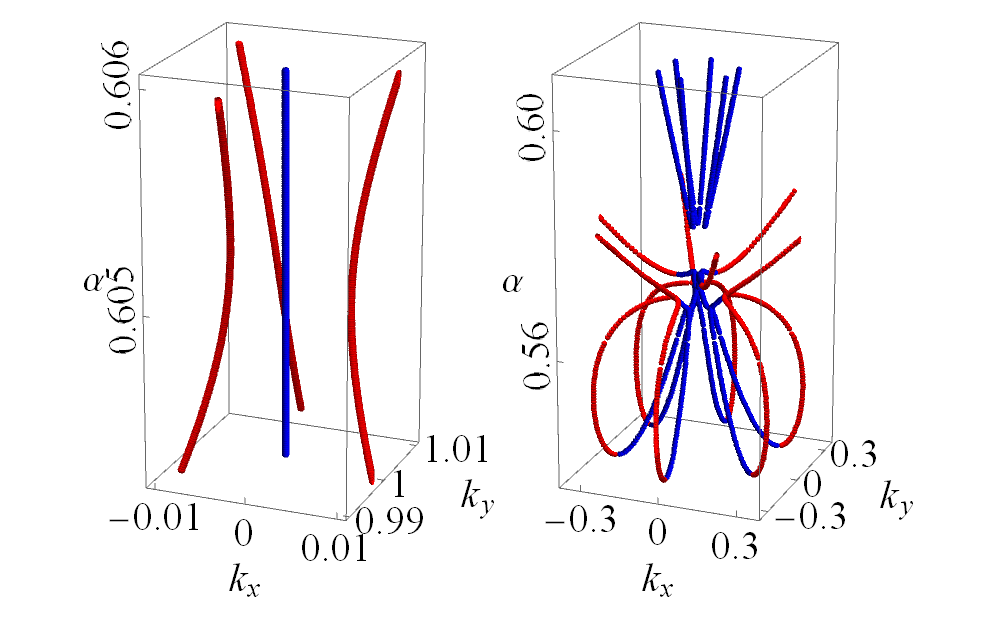}}
\\
\subfigure[]{\label{fig:nonBc}
\includegraphics[width=0.38\textwidth]{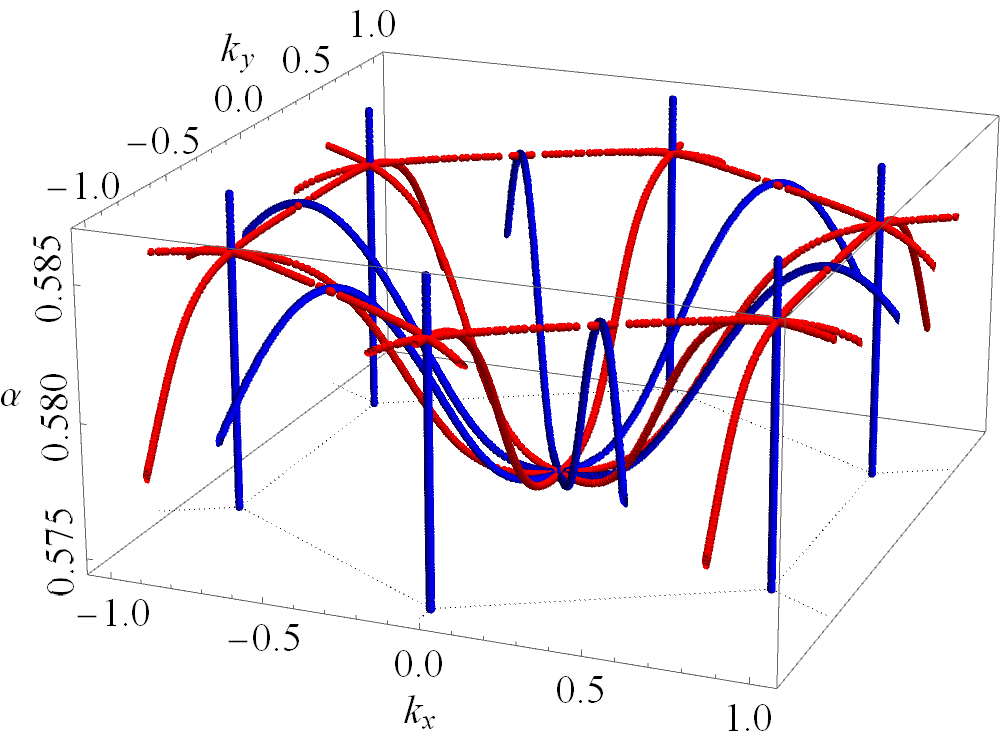}}
\hspace{0.2in}
\subfigure[]{\label{fig:nonBd}
\includegraphics[width=0.38\textwidth]{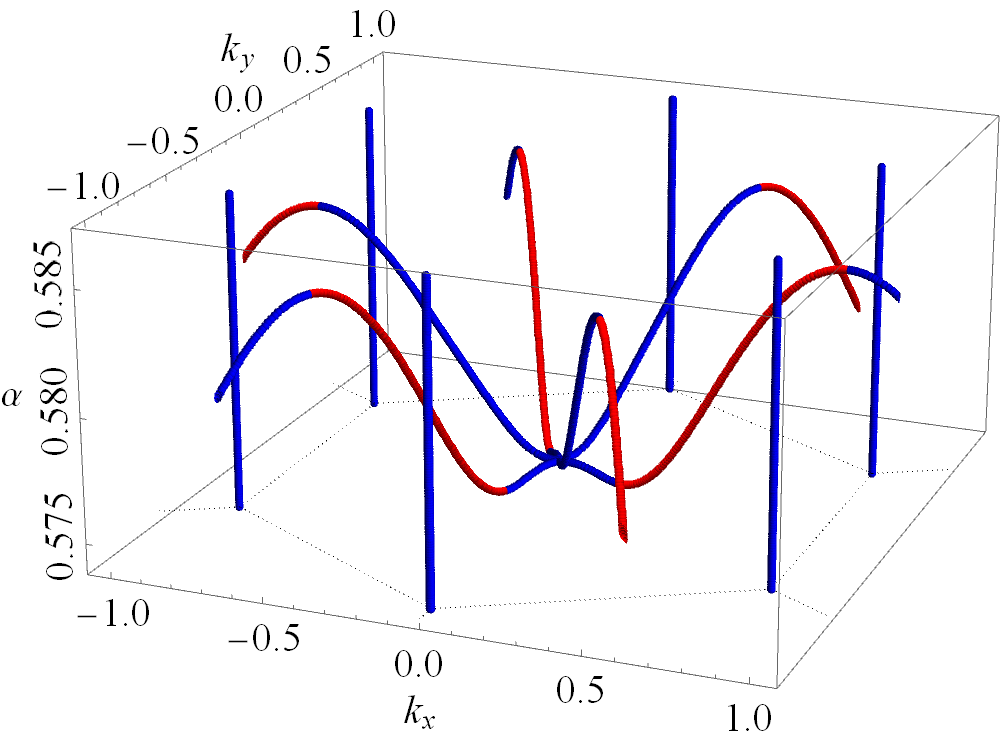}}
\caption{The topological transitions in the magnetic range for (a) the $(\eta,\mathbf{R}) = (1.0,\mathbf{R}_{\tau^z\theta/2})$ model, (c) the $(\eta,\mathbf{R}) = (0.82,1_{2\times2})$ model, and (d) the $(\eta,\mathbf{R}) = (0.82,\mathbf{R}_{\tau^z\theta/2})$ model. (b) shows the magnified version of the DP movements in (a) in the vicinity of $\mathrm{K}$ or $\mathrm{K}'$ (left) and also the $\Gamma$ point (right).}\label{nonB}
\end{figure}

In Ref.~\onlinecite{hejazi2019multiple}, the series of topological transitions near the magic angle for the case $(\eta,\mathbf{R}) = (1,1_{2\times 2})$ have been studied in detail. Here we will consider three other cases: 1). $(\eta,\mathbf{R}) = (1,\mathbf{R}_{\tau^z\theta/2})$, 2). $(\eta,\mathbf{R}) = (0.82, 1_{2\times 2})$, and 3). $(\theta,\mathbf{R}) = (0.82, \mathbf{R}_{\tau^z\theta/2})$. Note that the magnetic model considered in this work is based on the parameters of case 2). The DP evolution diagrams for these three cases as $\alpha$ is varied near the magic angle range are shown in Fig.~\ref{nonB}.

%Here we consider the topological transition happening in the non-magnetic continuum model around the first magic angle. We will consider three cases: 1. $\eta=1$ and sublattice pseudospin rotation due to twist angle considered; 2. $\eta=0.82$ and sublattice pseudospin rotation due to twist angle \emph{not} considered, and 3. $\eta=0.82$ and sublattice pseudospin rotation due to twist angle considered. The magnetic model considered in this work is based on the second case. Note that the sublattice pseudospin rotation introduces a small particle-hole asymmetry into the model.

Let us recall what happens in the case of $(\eta,\mathbf{R})=(\eta,1_{2\times2})$, considered in Ref.~\onlinecite{hejazi2019multiple}. When the twist angle is varied near the first magic angle, the topological transitions can be divided into two disconnected processes: first 6 pairs of DPs (each pair contains opposite chirality) emerge in the vicinity of $\Gamma$, and then pair-wise annihilate each other, again in the vicinity of $\Gamma$. The second part starts by the emergence of another six pairs of DPs at $\Gamma$, which later form six outer ones with the same chirality and six inner ones with the opposite chirality. The six inner ones participate in a further transition at the $\Gamma$ point, while the six outer ones pair-wise meet at $\mathrm{M}$ points, deflected towards $\mathrm{K}$ and $\mathrm{K}'$ points at which they participate in the Lifshitz transition. Finally, the six outer ones coming out from $\mathrm{K}$ and $\mathrm{K}'$ annihilate with the six inner ones coming out of the $\Gamma$ point. 

First, it is interesting to compare the topological transitions happening in the $(\eta,\mathbf{R})=(1,\mathbf{R}_{\tau^z\theta/2})$ case with the case of $(\eta,\mathbf{R})=(1,1_{2\times2})$ considered in Ref.~\onlinecite{hejazi2019multiple}. It turns out the topological transitions of the two cases resemble each other, in the sense that in both cases, DPs emerge near the $\Gamma$ point, and later on half of the DPs move towards $\mathrm{M}$, $\mathrm{K}$ and $\mathrm{K}'$, which finally annihilate with the other half. However, several important changes are observed immediately: 
\begin{itemize}
\item There are substantially larger number of topological transitions as the twist angle is varied near the magnetic angle. The two disconnected processes of topological transitions mentioned earlier for the $(\eta,1_{2\times2})$ model are now connected.
\item There are band touching points between active bands and their immediate higher neighbor band near $\alpha = 0.5724$, at which three pairs of DPs -- all six have the same chirality -- are annihilated in the vicinity of $\Gamma$ point, where their charges are transferred into higher bands.
\item Due to breaking of particle-hole symmetry, the topological transition originally present at $\mathrm{M}$ for the $(\eta=1,1_{2\times 2})$ model no longer takes place at $\mathrm{M}$, but in the \emph{vicinity} of it.
\item Importantly, there is \emph{no} Lifshitz transitions at or near $\mathrm{K}$ or $\mathrm{K}'$ points: as three pairs of DPs are deflected from the vicinity of $\mathrm{M}$, they induce a trigonal warping near $\mathrm{K}$ and $\mathrm{K}'$, but they no more pass through $\mathrm{K}$ or $\mathrm{K}'$; nor do they meet or get deflected near $\mathrm{K}$ or $\mathrm{K}'$. Instead, they come close to $\mathrm{K}$ or $\mathrm{K}'$, rotate $\pi/3$ around those two points, and leave towards $\Gamma$. They eventually annihilate with the other six DPs coming from the $\Gamma$ point. The Dirac velocity at $\text{K}$ and $\text{K}'$ does not vanish for any value of $\alpha$.
\end{itemize}
The last point mentioned above is particularly worth further discussion: the absence of topological transitions at (or near) $\mathrm{K}$ or $\mathrm{K}'$ means that no magic angles can be defined in the conventional sense; instead there is a ``magic range'' of twist angles where the active two bands become considerably flat. The absence of topological transitions at $\mathrm{K}$ or $\mathrm{K}'$ is a direct consequence of introducing the sublattice pseudospin rotation $\mathbf{R}_{\tau^z\theta/2}$. This term breaks the particle-hole symmetry, and allows a new term $\upsilon \bm{k}\times \bm{\sigma}$ into the low energy effective $\bm{k} \cdot \bm{p}$ Hamiltonian near $\mathrm{K}$ or $\mathrm{K}'$. Such an effective model is then written as
\begin{equation}
H_{\mathrm{K}/\mathrm{K}',\text{eff}} = v k^+\sigma^- + v^* k^-\sigma^+ + a k^{+2}\sigma^++ a^* k^{-2}\sigma^-,
\end{equation}
where $v = v_F + i \upsilon$ is a complex paramter; note that in the particle-hole symmetric case, $v$ is real and equal to $v_F=s(\alpha-\alpha_0)+O(\alpha-\alpha_0)^2$, i.e. the Dirac velocity is first order in $\alpha - \alpha_0$ near the first magic angle. The imaginary part, $\upsilon$, on the other hand is expected to be a constant at the leading order of $\alpha-\alpha_0$; however, since it vanishes when the particle hole symmetry is retained, it has to be linear in the twist angle $\theta$ which is a small parameter itself, i.e.~it should have a form $c_0 \theta$, with $c_0$ a constant. Defining $\varphi = \mathrm{arg}(v)$, the Dirac points surrounding $\mathrm{K}$ or $\mathrm{K}'$ are located at (in polar coordinates, with the pole set at $\mathrm{K}$ or $\mathrm{K}'$)
\begin{equation}
(k,\theta) = \left(\frac{1}{a}\sqrt{s^2(\alpha-\alpha_0)^2+ \upsilon^2},-\frac{1}{3} (\mathrm{arg}(a)+\varphi) + \frac{2\pi}{3} n\right),\quad n = 0,1,2.
\end{equation}
We see that $\upsilon$ prevents the three nearby Dirac points from visiting $\mathrm{K}/\mathrm{K}'$ and thus the Dirac velocity does not vanish.

%The complete topological transition process is summarize in Fig.~\ref{topological_transition_eta_1_thetadev_1}.

%\begin{figure}
%\centering
%\includegraphics[width=0.5\textwidth]{topological_transition_eta_1_thetadev_1.png}
%\caption{Complete topological transitions of the $(\eta=1,\mathrm{R}_{\tau^z\theta/2})$ model in the magic range. The additional ones compared with the $(\eta=1,1_{2\times2})$ case are highlighted in red.}\label{topological_transition_eta_1_thetadev_1}
%\end{figure}

Next we consider the topological transitions in the case of $(\eta,\mathbf{R}) = (0.82, 1_{2\times 2})$. The number of topological transitions is reduced down to three: first, 12 DPs (with alternating chiralities) emerge at the $\Gamma$ point, then, the six with the same chirality move towards $\mathrm{M}$ point and the other six with opposite chirality move towards $\mathrm{K}$ and $\mathrm{K}'$. The latter six then participate in the Lifshitz transition at $\mathrm{K}$ and $\mathrm{K}'$ in the same manner that one encounters in the $(\eta,\mathbf{R}) = (1.0,1_{2\times 2})$ model. Finally, these six DPs leave $\mathrm{K}$ and $\mathrm{K}'$ points and meet with the other six DPs exactly at $\mathrm{M}$ (where at each $\mathrm{M}$ point there are in total four DPs with total zero chirality), at which four DPs annihilate each other, leaving no further topological transitions behind. Note that the topological transitions are ``inverted'' in this case compared with the $\eta=1$ cases, in the sense that now the transition at $\mathrm{K}$ and $\mathrm{K}'$ precedes the ones at $\mathrm{M}$. This ``inversion'' signifies a transition behavior in the topological transitions themselves as $\eta$ is reduced from $\eta=1$ to $\eta=0.82$. This transition behavior is interesting in its own right, but we will not elaborate on it.

Finally we study the topological transitions in the case of $(\eta,\mathbf{R})=(0.82,\mathbf{R}_{\tau^z\theta/2})$. This is the most realistic of the three versions we considered. The topological transitions in the magic range is nevertheless quite simple. First, at $\alpha =0.574827$, three pairs of DPs emerge near $\Gamma$; the three DPs with negative chirality move towards $\Gamma$, at which point three new pairs of DPs are also created in quadratic band touching transition; the positively charged DPs annihilate the previous negative DPs and the new negative DPs  continue moving out in the same direction as the initial DPs; the three initial DPs with positive chirality move in the opposite way, travel across the BZ border, and finally pairwise annihilate with the ones with negative chirality. No topological transitions appear near $\mathrm{K}$ and $\mathrm{K}'$ points.

\section{Chern number}\label{app:chern}

Quite often the quantized Hall conductivity (in units of $e^2/h$) in an energy gap can be determined algebraically by writing down the Diophantine's equation describing the gap and then applying Streda's formula. The Chern number for a group of bands can then be obtained by computing the quantized Hall conductivity at the energy gap above and below this group of bands, and then take the difference. An equally good method to determine Chern number for a group of bands is to make direct use of the band topology by computing the berry curvature numerically. In this method, care must be taken to make sure the numerically computed berry curvature is gauge invariant as required by definition. Such a method has been developed in Ref.~\onlinecite{Fukui2005}, which applies to any Bloch Hamiltonian $H(\bm{k})$ which is periodic along the two directions of the (2D) Brillouin zone. The two methods have been shown to give identical results in several cases \cite{Aidelsburger2016,PhysRevB.74.205414}.

The situation, however, is subtle for the magnetic model (see App.~\ref{app:the_model}) we are using here: the hermitian matrix corresponding to the Hamiltonian in the Landau level basis, $H(\bm{k})$, does not have the ``expected'' periodicity, that is being periodic with respect to the reciprocal vectors of the magnetic translation lattice (see Eq.~\ref{eq:magneticBZ}).
%This is reminiscent of the standard problem of Bloch electrons moving in perpendicular magnetic field, considered in the TKNN paper \cite{thouless1982quantized}: there, the hermitian Hamiltonian matrix in the Landau level basis is also not periodic function of the BZ\footnote{
%	\blu{In Ref.~\onlinecite{thouless1982quantized}, the magnetic brillouin zone has a size of $[0,2\pi/qa]\times[0,2\pi/b]$. The Hamiltonian in the Landau level basis becomes a hermitian matrix $H(\bm{k})$. Translating $H(\bm{k})$ along the $k_x$ and $k_y$ period, one has $H(k_x,k_y+2\pi/b) = \Xi^{-1}H(\bm{k})\Xi$ and $H(k_x+2\pi/qa,k_y) = \Lambda^{-1}_p H(\bm{k})\Lambda_p$, where $\Xi$ is some permutation matrix and $\Lambda_p$ is some diagonal unitary matrix. Clearly $H(\bm{k})$ does not respect the $k_x$ periodicity of the BZ due to the permutation matrix $\Xi$; however the full Hamiltonian (involving the Landau levels) respect this periodicity.}
%}. 
The issue has to do with the fact that the Landau level basis we are using satisfies a \emph{generalized} Bloch theorem\cite{thouless1982quantized} rather than the original Bloch theorem. Consequently, the numerical method for computing the Chern numbers mentioned above has to be modified.

To give the modified numerical method, let us first establish the correspondence between the magnetic model in App.~\ref{app:the_model} and the standard problem of Bloch electrons moving in perpendicular magnetic field\cite{thouless1982quantized}. We first establish the correspondence between the Landau basis. Define $k_1 = y_0/\ell_{\text{B}}^2$, the Landau level basis used in App.~\ref{app:the_model} can be written in real space
\begin{equation}
\begin{aligned}
\langle \bm{x}|\tau,n,\sigma,y_0,j,k_2\rangle
&=\Gamma_{\tau,n,\sigma,j,\bm{k}}(\bm{x})\\
&=\sum\limits_{m=-\infty}^\infty
e^{-ik_2(x-(m+\frac{j}{q})pa)}e^{i\frac{2\pi}{a}(mq+j)y}\phi_n\left(y-\left(\left[m+\frac{j}{q}\right]pb-\frac{p}{q}\frac{ab}{2\pi}k_1\right)\right),
\end{aligned}
\end{equation}
where we defined lengths along $\hat x$ and $\hat y$ direction, respectively,
$$a = \frac{4\pi}{\sqrt{3}k_\theta},\qquad b = \frac{4\pi}{3k_\theta}.$$
This is essentially the same basis used in Ref.~\onlinecite{thouless1982quantized}, with a change of definition $(p,q)\rightarrow (q,p)$ and axis $(\hat{x},\hat{y})\rightarrow(\hat{y},\hat{x})$. The translation vector along $\hat{x}$ and $\hat{y}$ is then $a\hat{x}$ and $pb\hat{y}$; this essentially defines a rectangular magnetic BZ of size $2\mathcal{A}$:
$$(k_x,k_y) \in \left[0,\frac{2\pi}{a}\right]\times\left[0,\frac{2\pi}{pb}\right].$$

The Bloch function $|u(\bm{k})\rangle$ is then written as
$$|u_{\tau,\sigma}(\bm{k})\rangle = \sum_{j=0}^{q-1}\sum_{n=0}^\infty
d_{\tau,\sigma,n,j}|\tau,n,\sigma,y_0,j,k_2\rangle.$$

The Bloch Hamiltonian $H(\bm{k})$ constructed in App.~\ref{app:the_model} gives a Hermitian matrix in the basis $|\tau,n,\sigma,y_0,j,k_2\rangle$. One then diagonalizes this Hermitian matrix numerically to obtain the eigenvectors $\mathbf{d}(\bm{k})$, with entries $d_{\tau,\sigma, n,j}(\bm{k})$.

We now derive the modified numerical method to computing the correct band Chern number. Remember that in the original method of Ref.~\onlinecite{Fukui2005}, the band Chern number is obtained by summing over berry curvature of this band (or multiple bands) in the BZ; the berry curvature $\mathcal{B}(\bm{k})$ is computed numerically from the eigenvectors $\mathbf{d}(\bm{k})$ using a discretized and gauge invariant version of the definition:
\begin{equation}
\mathcal{B}_{\text{numerical}}(\bm{k})
=i\epsilon_{ij} 
\left(\partial_{k_i}\mathbf{d}(\bm{k})\right)^\dag\partial_{k_j}\mathbf{d}(\bm{k}),
\end{equation}
where $\epsilon_{ij}$ is the 2D antisymmetric tensor and $i$ and $j$ are implicitly summed.

To remedy this, one must remember that here we are using Landau levels as the basis; thus the basis states are also functions of momentum, and may as well contribute to berry curvature. The complete Berry curvature is then computed as follows:
\begin{equation}\label{eq:correctBerry}
\begin{aligned}
\mathcal{B}(\bm{k}) &= i \epsilon_{ij}\left(\partial_{k_i}\langle u(\bm{k})|\right)\partial_{k_j}|u_{\bm{k}}\rangle\\
&=i\epsilon_{ij} 
\left(\partial_{k_i}\mathbf{d}(\bm{k})\right)^\dag\partial_{k_j}\mathbf{d}(\bm{k})
+i\epsilon_{ij}\sum_{\tau,\tau',\sigma,\sigma',j,j',n,n'}d^*_{\tau,\sigma,n,j}(\bm{k})d_{\tau',\sigma',n',j'}(\bm{k}) \left(\partial_{k_i}\langle \tau,n,\sigma,y_0,j,k_2|\right)\partial_{k_j}|\tau',n',\sigma',y'_0,j',k'_2\rangle,
\end{aligned}
\end{equation}
note that the Berry curvature receives contribution from two parts, the first from the eigenvector $\mathbf{d}$ and the second from the LL basis $\Gamma$. 

Now we integrate both sides over the BZ to obtain the Chern number $C$. Note that the integral of the second term can be calculated analytically, which gives a $-1/q$ contribution. Therefore we have
\begin{equation}\label{eq:finalBerry}
C = C_{\text{numerical}}- 1/q,
\end{equation}
i.e. each LL contributes a fraction $1/q$ to the total Chern number. Note the second term in Eq.~\eqref{eq:correctBerry} does not contribute to Chern number if the Bloch Hamiltonian matrix is written under a plane wave or Wannier basis. 

%\sout{As an application of the formula \eqref{eq:finalBerry}, let us study the band topology of the model in App.~\ref{app:the_model}. The computed Chern numbers for several groups of bands at both small (left wing of the butterfly) and large (right wing of the butterfly) magnetic field limit are given in Fig.~\ref{fig:adiabaticity}. In the small magnetic field limit, the Chern numbers agree with the previous result obtained from Streda's formula \cite{PhysRevB.85.195458}. In the large magnetic field limit, however, the groups of magnetic bands all have vanishing Chern number. This result is violating both Streda formula and the Chern numbers predicted from the dual theory in App.~\ref{app:infinite_B}.}

As an application of the formula, let us study the correspondence between the Chern numbers obtained in the original magnetic theory (App.~\ref{app:the_model}) at large flux and that obtained in the dual theory at small flux (see App.~\ref{app:infinite_B}). The Chern number of the dual theory (the honeycomb Hofstadter butterfly problem, i.e. Eq. \eqref{eq:honeycombdual}) has been obtained previously\cite{PhysRevB.74.205414,Agazzi2014}: each band at the edge contributes Chern number $C=1$ and each band at the charge neutrality contributes $C=2$. In fact the relation between the Chern number sequence $2,2,...,1,1$ of the honeycomb Hofstadter butterfly model and the
 vanishing of Chern number of our model in App.~\ref{app:the_model} at large field can be understood using Eq.~\eqref{eq:correctBerry}. Take the group of $4p$ bands at charge neutrality of our model at large field as an example: since $C=0$, $C_{\text{numerical}} = 4p/q$. 
The dual theory has a dual magnetic BZ with an area
 that is equal to $q/2p$ times the MBZ area of the model in App.~\ref{app:the_model}. Since the Chern number for the dual magnetic model, $C_{\text{dual}}$, is evaluated on the dual magnetic BZ, and noting that the MBZ of the model in App.~\ref{app:the_model} is $p$-fold degenerate (consisting of $p$ identical segments along $k_x$ direction when computing Berry curvature), we have
\begin{equation}
C_{\text{dual}} = \frac{q}{2p} \cdot C_{\text{numerical}} = 2,
\end{equation}
this is exactly the Chern number for the zeorth Landau level of the magnetic honeycomb \cite{PhysRevB.74.205414,Agazzi2014}. Note the second term in Eq.~\eqref{eq:correctBerry} does not contribute to the Chern number of the dual theory, since the Hamiltonian is written in plane wave basis.

Using this line of argument, all Chern numbers in the original and the dual theory can be understood.

\end{document}